\documentclass[preprint,pre,groupedaddress]{revtex4}
\usepackage[pdftex]{graphicx,epsfig}

\begin{document}

\title{Escalation, timing and severity of insurgent and terrorist events: Toward a unified theory of future threats}

\author
{Neil F. Johnson}
\affiliation{Physics Department, University of Miami, Coral Gables, FL 33126, U.S.A.}

\date{\today}

\begin{abstract}
I present a unified discussion of several recently published results concerning the escalation, timing and severity of violent events in human conflicts and global terrorism, and set them in the wider context of real-world and cyber-based collective violence and illicit activity. I point out how the borders distinguishing between such activities are becoming increasingly blurred in practice -- from insurgency, terrorism, criminal gangs and cyberwars, through to the 2011 Arab Spring uprisings and London riots. I review the robust empirical patterns that have been found, and summarize a minimal mechanistic model which can explain these patterns. I also explain why this mechanistic approach, which is inspired by non-equilibrium statistical physics, fits naturally within the framework of recent ideas within the social science literature concerning analytical sociology. In passing, I flag the fundamental flaws in each of the recent critiques which have surfaced concerning the robustness of these results and the realism of the underlying model mechanisms. 
\\

\noindent {\em Working paper to accompany an upcoming seminar discussion. The paper will be updated sporadically as the research program to quantify and model collective human predation in real-world and cyberspace systems evolves.}
\end{abstract}

\maketitle

\section{THE COMPLEX SPATIOTEMPORAL DYNAMICS OF CONFLICT}
Irrespective of its origin, any given conflict or terrorist campaign will play out as a  highly complex dynamical system driven by interconnected actors whose actions are driven by a wide variety of evolving information sources, myriad socioeconomic, cultural and behavioral cues, and multiple feedback processes. Furthermore, since conflicts and campaigns have a beginning and eventually an end, they will by definition exhibit non-steady state, out-of-equilibrium dynamics. To see the complications that can arise from the inherent feedback and nontrivial temporal evolution characteristic of such complexity, one does not necessarily have to limit consideration to the recent conflicts in Iraq and Afghanistan, or recent global terror campaigns. In the current year alone (i.e. 2011) three examples have arisen in which a loose, quasi-coherent coordination between collections of humans has given rise to violent events in response to news or rumors circulating on the Internet or electronic messaging between participants: The Arab Spring uprising,  the London riots in the United Kingdom, and the high profile cyber-attacks on government and commercial sites by secretive hacker groups such as LulzSec. Indeed, the mixing of events in real and cyber space, together with the fueling of illicit activities by the drug trade and international crime, has led to a blurring of the boundaries between terrorism, insurgency, war, so-called organized crime, and common delinquency. For example, there is a clear and present threat to the U.S. from Mexico as a result of the combination of such activities: On September 2010, U.S. Secretary of State Hillary Clinton, in remarks to the Council on Foreign Relations, said that the violence by the DTOs (Drug Trafficking Organizations) in Mexico may be ``morphing into, or making common cause with, what we would call an insurgency" \cite{marines}. The United Nations, in its report titled ``The Globalization of crime: A transnational organized crime threat assessment" \cite{UN}, cites a statement by the UN Security Council from February 2010 in which they highlight ``.. the serious threat posed in some cases by drug trafficking and transnational organized crime to international security in different regions of the world". This report also states that a major stability threat will arise in the future as cartels and gangs fight for control of territory, and that the impact will be increasing violence among criminals and toward state officials and the public. In short, there will be more likelihood of spillover violence, more corruption, increasing loss of territorial control, and an accelerated undermining of both the social fabric and homeland security for nation states. 

Interrelated to the situation in Mexico, is that of Colombia, where a thirty-plus year war still continues. Though Marxist in origin, its character has been mixed up by the narcotraffic industry, criminal gangs, mafia cartels, paramilitary groups, the presence of at least two major guerilla organizations (FARC and ELN), and widespread common delinquency driven by a variety of socioeconomic factors \cite{bbcColombia}. As such, the struggle faced by state organizations to counteradapt to ever-changing guerilla-narco-crime-cartel innovations, is immense. Quoting Ref. \cite{bbcColombia}, President Santos recently outlined new tactics to counteradapt to the guerrillas adoption of (i) hit-and-run raids using flexible units, (ii) mixing of rebels and criminal gangs and their use of joint activities as mutual needs arise, for example so-called Bacrims  which are organized criminal bands, 
(iii) dressing of insurgents as civilians to merge into the general population, (iv) carrying
out small-scale attacks for maximum attention but little risk to themselves. 
In response to the dynamical and fragmented nature of the insurgent cells which threaten the government's control on law and order, the state's new tactics include more flexible units in order to mirror more closely the insurgents' behaviors: ``We have to adjust our doctrine, our operations and our procedures to the way (the rebels) are
operating" (President Juan Manuel Santos \cite{bbcColombia}). These features (i)-(iv) of an insurgent Red force are not unique to Colombia -- they reflect the behaviors likely adopted by {\em any} armed group on the Red side that is fighting to survive, whether it operate in real space or in the cyberworld, or some future hybrid mix of the two \cite{gambetta,robb,kenney}. For this reason, these properties (i)-(iv) will play a core mechanistic role in the generic model presented in this paper.
Indeed there is an entirely parallel threat to real-space insurgencies and terrorism which is evolving on the Internet, in terms of transnational attacks in the cyber domain from both sovereign state and non-state actors.  This threat is arguably even more urgent than the real-space one, given that cyber `weapons' (e.g. encounter-network worms or bots) can be assembled very quickly, and transported in principle at the speed of light (i.e. via communications limits within fiber-optic networks). The advantage for Red (i.e. an insurgent or illicit organization) is that these cyber-logistics are much easier, quicker, and naturally more clandestine than the physical task of having to transport weapons and/or people from a point of assembly to the place of potential attack. 

Future predatory threats in real and/or cyberspace, are likely to adapt to, and exploit, the rapid, ongoing advances in global connectivity, and hence present clear but evolving dangers to each and every nation state, corporation or legitimate organization. The resulting arms-race between adaptation-counteradaptation by Red and Blue (i.e. state organizations) will likely lead to rapid innovation of new predation methods. In addition, the background civilian population, refereed to here as Green, cannot be considered as purely passive -- instead, it is a three-way struggle between Red, Blue and Green. Given this complexity, the possibility for rapid escalation of hybrid real-world attacks, cyber attacks, and cyber-assisted attacks, therefore represents an unprecedented future risk which needs to be understood, quantified, mitigated and controlled -- or at least delayed or deflated in terms of its potential impact. But there are so many questions that need addressing: How are these national and international threats likely to evolve going forward? Given their finite resources, how can state agencies and countries be best prepared to face this challenge? What can be done in advance to prepare for the next generation of cyber--real-space attacks? Are there any likely points of intervention that can be usefully exploited? Without quantitative models of such situations, solutions must be sought purely on the basis of narratives and case-studies (assuming any are available). It is clear that such narratives and case studies could play a crucial role, in particular where very few prior examples are known, or where strong socioeconomic, cultural or behavioral factors play a key role. But as the amount of available data from such attacks, both in the real world and cyber world, increases, is there anything additional that can be said from a statistical viewpoint? In particular, given that human conflicts and terror campaigns are examples of a highly complex dynamical system driven by interconnected issues and actors, is this a potentially fruitful topic for analysis within the framework of the statistical physics of non-equilibrium open systems? And might such a study then in turn shed valuable empirical light back on the emerging field of non-equilibrium statistical physics?

In this paper, I review some recent steps taken in this direction by our collaborative team. The results in many ways build upon earlier works within the statistical physics community, e.g. Redner \cite{redner} and Rodgers \cite{rodgers} among others \cite{ez,opinion,McKane,galam,bursts,palla,necsi,ian,ian2}, going all the way back to the mid-20th century work of Lewis Fry Richardson \cite{richardson} and Frederick Lanchester \cite{lanchester}. In particular, we pursue a methodology which complements that of the political, social and life science fields \cite{kilcullen,politics,larserik1,larserik2,larserik3,larserik4,larserik5,larserik6,larserik7,larserik8,larserik9,larserik10,
larserik11,larserik12,dom1,dom2,dom3,dom4,dom5,john,gambetta,robb,kenney,
caro,horgan1,horgan2,horgan3,horgan4,horgan5,horgan6,horgan7,horgan8,horgan9,horgan10,horgan11,carley1,carley2,carley3,carley4,
carley5,carley6,carley7,carley8,carley9,carley10,carley11,carley12,carley13}, but which mirrors the approach initiated within the statistical physics community in association with the study of financial markets \cite{Bouchaud,Stanley,econophysics,book}. Our methodological program follows these steps: (1) Use state-of-the-art spatiotemporal datasets with the highest available resolution, combined with current narratives from the academic literature, online sources, and the broader national and international media, in order to identify systematic and anomalous behaviors in the ongoing timelines of daily, weekly and monthly events within a given domain of human predation. (2) Quantify the resulting stylized statistical facts of these multi-component time-series and hence identify statistically significant deviations or anomalies. (3) Carry out a parallel procedure for other predation domains (e.g. provinces, or countries, or cultures) identifying where and when similar stylized facts emerge and, by contrast, where anomalies arise. (4) Develop a multi-agent mechanistic model of the underlying multi-actor dynamics for the domains of interest. Then undertake an iterative process of model modification and comparison to the data in order to obtain a minimal mechanistic model which is consistent with the most robust stylized facts which have been extracted from the data. 

The individual works which I summarize here using a unified perspective, have appeared separately in a range of different specialized journals \cite{science,nature,website,duration,blazej,aip,gangs,wars,clustering,ezsir,newsimpact,police,aps,conflict2,bias1,bias2,predict,predict2,manybody,emg} following our two initial preprints \cite{lanl1,lanl2} in which we presented an initial coalescence-fragmentation model and an initial analysis of the empirical data for the severity of events. Many others have presented excellent complementary or related works \cite{redner,rodgers,galam,Lauren}, but I do not discuss these works here or give a comprehensive review. Despite this progress, much remains to be done. In particular, our ongoing projects are focused around connecting the above listed works to the wider body of research activity within the human social, cultural and behavioral domain \cite{horgan1,horgan2,horgan3,horgan4,horgan5,horgan6,horgan7,horgan8,horgan9,horgan10,horgan11,carley1,carley2,carley3,carley4,carley5,carley6,carley7,carley8,carley9,carley10,carley11,carley12,carley13}. Examples that we are pursuing include adapting the model from step (4) above, to incorporate the results from studies from social psychology \cite{mike}, economics, social science and crime science \cite{felson}. These results from other fields, while quantitative, may not be in the form of high frequency time-series, the preferred choice of the statistical physics community \cite{econophysics} -- hence their incorporation represents a challenge to the statistical physics modeling process. However their inclusion is essential in order for such mechanistic analysis to be taken seriously by the other communities studying collective human predation. We are also actively probing time-series anomalies (e.g. see Red Queen discussion below) and interpreting them in terms of actor decisions, adaptations and counteradaptations. We are also using the multi-agent model to interpret how the underlying armed actors are themselves adapting their strategies, changing their membership and reach, and counteradapting to current government and agency procedures. In parallel to earlier work in econophysics, we can then run these scientific models forward in time in to order to develop realistic prediction corridors for where each of the indicators are likely to evolve and hence how the future threat is likely to change \cite{predict,predict2}. This will allow us to change features in the parameter space of the models, in order to test out scenarios and identify points in the future evolution at which interventions might be possible, or necessary, in order to avoid particular undesirable outcomes \cite{predict,predict2}.

\begin{figure}
\includegraphics[width=0.95\linewidth]{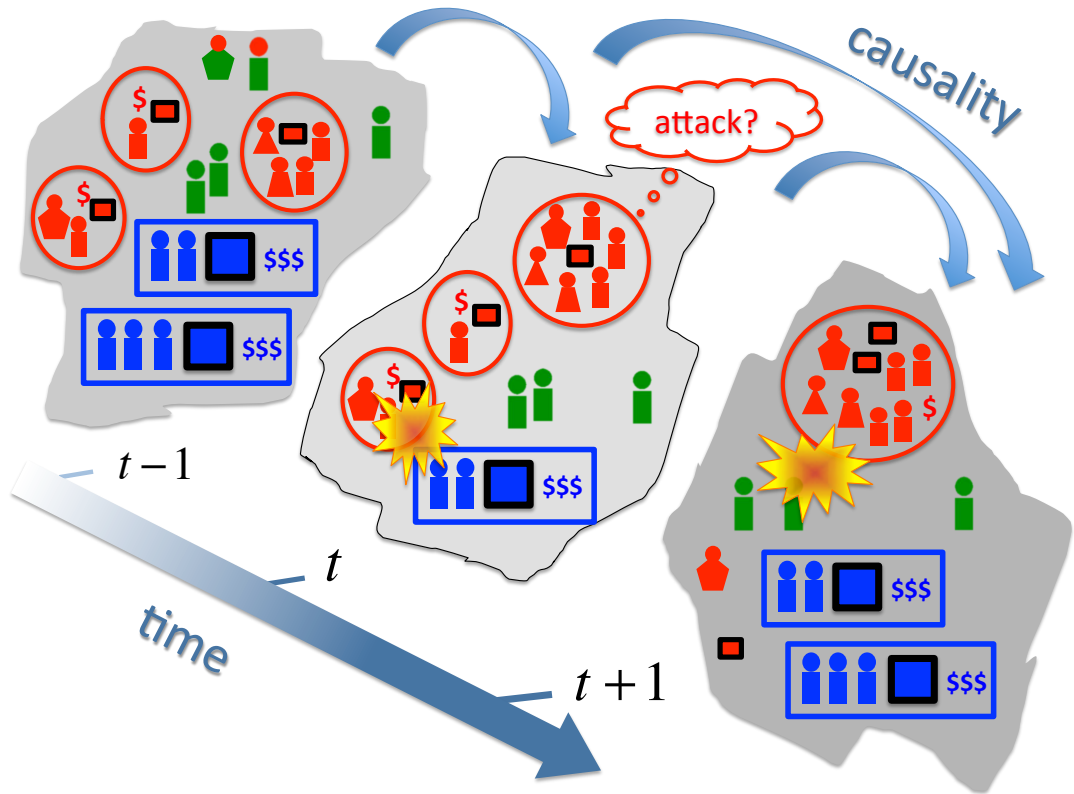}
\caption{(Color online) Schematic of the complex spatiotemporal dynamics of multi-actor collective predation in real and/or cyber space. The result is a complex ecology of interactions and observed events, driven by some dynamically evolving but hidden network of loosely connected Red cells featuring non-local interactions aided by electronic communications \cite{lanl1,lanl2,nature}. At any one time, there may be multiple types of actor, and these may cross different cultural and behavioral boundaries. Each population is partitioned into loose temporal cells \cite{bbcColombia,gambetta,robb,kenney}. Each cell may itself  sporadically combine with another cell, or simply fragment in some way -- for example, as a result of sensing danger \cite{gambetta,robb,kenney,caro,ian,ian2}. In addition to the traditional Blue population  (e.g. state military, terrorist group or intelligence organization) and Red (e.g. insurgency or hacker group), there is also a background civilian population which is labelled as Green, but which may not be passive in the struggle. Social, cultural and human behavioral factors may play an important role for Green and Red at the level of individual cells and its members.}
\label{fig1}
\end{figure}

Even the simple cartoon representation in Fig. 1 demonstrates that at any one timestep, the complexity of the actors and their interactions can create a formidably complicated dynamical system. For studies of fatalities, the observable output ${\bf x}_i(t)$ can be considered a vector whose elements describe the number of fatalities for each population type (i.e. Red, Blue, Green) at place $i$ at timestep $t$. More generally, the output ${\bf x}_i(t)$ would be a tensor, showing separately the numbers of victims killed and wounded, and the different weapon types used (e.g. Improvised Explosive Device (IED), or suicide bomb, or rocket propelled grenade (RPG), or small arms fire). For simplicity, we will tend to refer to the `Red' population as `insurgent', even though they may be a heterogeneous collection of cyber-gangs, drug cartels, idealistic insurgents, rebels or rioters, and we refer to `Blue' as the `coalition military' or `official antiterrorist organization' even though they may be cyber-defense, police, security forces etc. Setting aside the issue of whether the data recorded has an observational bias or not due to the way it was recorded (e.g. main street bias \cite{bias1,bias2}), there are many other potential complications facing a data-driven research program such as ours. These include, but are not limited to, the following: (i) Heterogeneity of the insurgent force strength (i.e. Red) which is depicted in Fig. 1 as various `types' of fighter, or weapons, or assets including financing. This could also include different cultural, social and behavioral types within Red. Even the assumption that there is just one Red force can be misleading, as evidenced currently in Colombia (ELN, and FARC) and in Libya, with the different rebel factions. It is not just an `us and them' situation. (ii) Heterogeneity of Blue, comprising warfighters, equipment and money. (iii) Heterogeneity of Green, the background civilian population, in terms of tribal or ethnic groups. (iv) The non-passive nature of Green due to possible influence, sympathy, or direct recruitment to Red. For example in Fig. 1, active support of Red is indicated by two green figures with red heads who then get converted in the next timestep to Red. Or it could simply be that a Green member shows an active failure to support Blue. (v) Changing number of Red members, or Red cells. (vi) Finite lifetime of any given Red cell due to endogenous or exogenous factors, such as its implicit fragility in the presence of Blue or when perceiving imminent detection or capture. The grouping dynamics that occur within and between insurgent and terrorist cells, and other illicit group activities, are unlikely to be of the form seen in more open social settings. As stated by Diego Gambetta in his influential book `Codes of the Underworld', on p. 5. , ``.... contrary to widespread belief, criminal groups are unstable \cite{gambetta}. In the underworld, people have a higher rate of mobility (and mortality) than most professions." This is also supported in the case of insurgencies by accounts such as by Robb and Kenney \cite{robb,kenney}. Such fragmentation under danger is also entirely consistent with observed antipredator defenses in birds and mammals \cite{ian2,caro}. 
(vii) Decisions by Red cells to attack are not made in isolation, nor are they irrespective of the past. Instead there is a complex, possibly unknowable, mix of past events which affect a given cell or its members in particular ways -- just as it does in the non-violent world of collective human struggles, e.g. financial market predatory trading \cite{book,wang}. In addition there is the convoluted effect that current and past exogenous and endogenous events and news might have, as is also known from the predatory environment of financial markets \cite{newsimpact}. These reactions to past and present events will also likely depend nonlinearly on social, cultural and behavioral factors. (viii) The nature of the observable events themselves: Even if they are accurately recorded, complete information will never be known precisely about who did what and why. For these reasons, the challenges facing anyone such as ourselves who wishes to analyze high-resolution spatiotemporal datasets recording the results of collective human violence, and then look for common stylized facts, and then finally build minimal mechanistic models, are highly nontrivial. Indeed, the fact that more detailed spatiotemporal data is now becoming available, often down to the daily scale within individual provinces or districts, means that the bar has been raised in terms of what a model needs to achieve in order to be considered `consistent' with the data.

It is only recently that detailed attempts to compare mechanistic conflict models to datasets of day-to-day casualties, have appeared in the academic literature. This is in large part due to the fact that large, high-frequency datasets have only recently become available beyond confidential military circles, e.g. www.icasualties.org which gives coalition casualties in Afghanistan and Iraq. However, theoretical attempts to model human conflict mathematically have had a long history. They tend to break down into one of two extremes: Models which resemble chemical reactions, either in the form of continuous differential equations (e.g. Lanchester \cite{lanchester,niall1,niall2,niall3,niall4}), and computationally intensive individual-based models on some kind of fixed grid such as a checker-board or static spatial network \cite{Lauren}. In contrast to the situation many decades or centuries ago, however, there are a number of additional complicating factors: First, the classic image of a battle being fought between two well-regimented armies lining up at dawn on opposite sides of a field or plain, does not describe the fragmented, fluid situation of modern insurgencies \cite{gambetta,robb,kenney}, either in the real or cyber worlds. Second, broadcasting communications now exist in which events and images can be portrayed almost instantly to a broad sector of the global population, thereby possibly influencing the decisions of their elected leaders and respective security forces. Third, personal media resources such as Twitter and Facebook, together with texts and emails, mean that fighters (and potential fighters) who are separated across different streets, or towns, or countries, or continents, can be connected together within a second -- and hence they can coordinate their actions such that they begin to behave as one quasi-coherent group (or `cell'), even though they may never have met each other and may even be on different continents. It can also happen that the members of such a cell -- who may not be physically connected, but whose actions are somehow coordinated through the use of technology -- suddenly lose their collective coherence (e.g. loss of communications, or loss of trust) and hence the cell has effectively fragmented. At the touch of a keystroke or press of a button on a cellphone keyboard, they instantaneously disappear into the background noise generated by everyday human activities. Fourth, the distinction between an insurgent or terrorist (i.e. Red) and the background civilian population (i.e. Green) can be blurred and itself highly fluid. It is no longer the case that a civilian population can be considered some inert background which simply soaks up the violent events as they play out. We note that in addition to such mechanistic descriptions, the literature also contains statistical studies of total casualty count distributions inspired by the work of Richardson \cite{richardson}. For example, we \cite{lanl1,lanl2,nature}  have analyzed the distribution of events in insurgent wars and terrorism, while others have looked at terrorism \cite{clauset1}.

Our own approach to modeling focuses on the underlying ecology, by representing the various actors as interacting populations of heterogenous agents who operate with covert but dynamically evolving communication networks, and who adapt their strategies in response to external events and news, as well as counteradaptation by the relevant state authorities. Our overall vision of the complex global interaction between gangs, cartels, illicit crime groups etc. is therefore that of a complex ecology whose dynamics and internal interactions may change and adapt over time, with heterogeneous actors, interactions over space and time, adaptation-counteradaptation, feedback, and movement or communication via some underlying dynamical network. This view is in accordance with the state-of-the-art view of modern violent gangs proposed by Felson \cite{felson}, and the descriptions of Kilcullen, Robb and Kenney \cite{gambetta,robb,kenney}. Our mechanistic methodology is also entirely consistent with the analytical sociology viewpoint which is current in the social sciences \cite{hedstrom}. Adopting this viewpoint, we have managed to uncover 
common patterns in 
insurgent wars and global terrorism \cite{science,nature}, as well as a mechanistic relationship with the dynamics of street gangs and online guilds in massively multiplayer online cyberwar games \cite{gangs}.  In addition to numerical results, we have been able to obtain quantitative insight through the development of mathematical analysis which reproduces the main features of these numerical and statistical results \cite{duration,blazej,aip,wars}.
Our explanation for the existence of these common features, is simply that there are common ways in which humans `do' groups and group activities -- just as traffic jams can exhibit common statistical and dynamical properties in major cities around the world, and stock markets exhibit common stylized statistical facts and dynamics \cite{Bouchaud,Stanley,book}. As discussed later in reference to Fig. 8, our model of collective human predation is consistent with several recent hypotheses concerning modern insurgency \cite{kilcullen,robb}, is robust to many generalizations \cite{blazej,wars}, and establishes a quantitative connection between human insurgency, global terrorism and ecology. Its similarity to financial market models\cite{rodgers,ez} provides a surprising link between violent and non-violent forms of human behavior. At the operational level, our findings have a number of implications -- for example, the clustering dynamics imply that operational cells on the ground effectively have no permanent hierarchy or leaders. Our work also allows us to explore `what if?' scenarios in search of black swan behaviors and beyond.
Our data sources are real-time media databases, official (government and non-governmental organization) reports, and academic studies. The individual papers referenced provide full details. 

Before proceeding, we note that there is a possible confusion in terms of terminology regarding what to call clusters of insurgents etc. Just as any other social setting, a small cluster of people may be called a group, a team, a cell -- likewise a larger cluster may also be called a group, a crowd, or even an organization. Similarly, terrorists and insurgencies are sometimes referred to as `groups' even though this could be the entire entity (e.g. all members of the FARC and their infrastructure) or just a few members who happened to be together on a particular attack. In order to avoid a misunderstanding of what constitutes a group, a cell, and an organization, we will try to adopt the language in which a cell is a cluster of a few Red agents (e.g. insurgents) which carries out a given attack, and organization is the entire Red outfit -- even though we stress that we do not want to assign any specific organizational capabilities, or assume that Red is necessarily well organized, or following a hierarchy. Indeed, as we will show, one of the implications of our work is that the cells are loose and transient in terms of their operational activity. This is one of the reasons they are probably so hard to track, in both real and cyber space.

\section{THEORETICAL BACKGROUND}
Though inspired by work in non-equilibrium statistical mechanics, it turns out that our mechanistic approach is remarkably consistent with current thinking in the social sciences Ð in particular, analytical sociology as developed by Hedstrom \cite{hedstrom}. In particular, Hedstrom states \cite{hedstrom} ``The basic idea of a mechanism-based explanation is quite simple: At its core, it implies that proper explanations should detail the cogs and wheels of the causal process through which the outcome to be explained was brought aboutÉ.. Mechanisms consist of entities (with their properties) and the activities that these entities engage in, either by themselves or in concert with other entities. These activities bring about change, and the type of change brought about depends on the properties of the entities and how the entities are organized spatially and temporally." Hedstrom goes on to state that the ``key challenge is to account for collective phenomena that are not definable by reference to any single member of the collectivity. Among such properties are 1. Typical actions, beliefs, or desires among the members of society or a collectivity. 2. Distributions and aggregate patterns such as spatial distributions and inequalities. 3. Topologies of networks that describe relationships between members of a collectivity. 4. Informal rules or social norms that constrain the actions of the members of a collectivity." Paraphrasing Hedstrom \cite{hedstrom}, a basic point of the mechanism perspective is that explanations that simply relate macro-properties to each other are unsatisfactory. He goes on to state that these explanations do not specify the causal mechanisms by which macro-properties are related to each other. It seems that deeper explanatory understanding requires opening up the black box and finding the causal mechanisms that have generated the macro-level observation \cite{hedstrom}. According to Hedstrom, social mechanisms and mechanism-based explanations have, over the past decade, received considerable attention in the social sciences as well as in the philosophy of science. As also stated by Hedstrom, some writers have described this as a mechanism movement that is sweeping the social sciences \cite{norkus}. He gives the example of a car's engine whose mechanisms and parts are quite visible when the hood is opened \cite{hedstrom}. Hedstrom also states that ``when one appeals to mechanisms to make sense of statistical associations, one is referring to things that are not visible in the data, but this is different from them being unobservable in principle".

Predator-prey systems have themselves been widely studied by many disciplines, including physics \cite{McKane}. Outside the few-particle limit, mean-field mass action equations such as Lotka-Volterra can provide a fair description of the average and steady-state behavior, i.e. $dN_R(t)/dt=f(N_R(t),N_B(t))$ and $dN_B(t)/dt=g(N_R(t),N_B(t))$ where $N_R(t)$ and $N_B(t)$ are the Red and Blue population's strength at time $t$. However, such population-level descriptions of living systems do not explicitly account for the well-known phenomenon of intra-population group (e.g. cluster) formation \cite{caro}, leading to intense debate concerning the best choice of functional response terms for $f(N_R(t),N_B(t))$ and $g(N_R(t),N_B(t))$ in order to partially mimic such effects. Analogous mass-action equations have been used to model the interesting non-equilibrium process of attrition (i.e. reduction in population size) as a result of competition or conflict between two predator populations in colonies of ants, chimpanzees, birds, Internet worms, commercial companies and humans in the absence of replenishment. The term attrition just means that `beaten' objects become inert (i.e. they stop being involved), not that they are necessarily destroyed. 
The combined effects of intra-population grouping dynamics and inter-population attrition dynamics have received surprisingly little attention \cite{redner,caro}, despite the fact that grouping and attrition are so widespread \cite{caro} and the fact that their coexisting dynamics generate an intriguing non-equilibrium many body problem.

\section{ESCALATION: PROGRESS CURVES AND THE DYNAMICAL RED QUEEN MODEL}
Before moving on to discuss specific mechanistic interactions at the level of individual cells, and hence necessarily introducing some assumptions concerning how they operate, I will start by taking a broad-brush view of the overall arms-race struggle between Red and Blue. This work is given in detail in Ref. \cite{science}.

We consider Red (e.g. insurgents) as continually wishing to damage Blue (e.g. kill coalition military). All other things being equal, Red would like to complete successful attacks as quickly as possible so that successive successful attacks become more frequent. Using coalition military fatality data for Afghanistan, we therefore analyzed the times for successive fatal days for the coalition military, and find that they follow an approximate power-law Ôprogress curveÕ $\tau_n=\tau_1 n^{-b}$ \cite{science}. Here $\tau_n$ is the time between the $(n-1)$Õth and $n$Õth fatal day, $\tau_1$ is the time between the first two fatal days, and $b$ controls the escalation process. A fatal day is one in which the insurgent activity produces at least one death. In particular, we calculated the best-fit power-law progress curve parameters $b$ and $\tau_1$ for each province. Figure 2 shows a remarkable linear relationship which then emerges between these best-fit progress curve values for different provinces. Examples of the best-fit progress curves are given in Fig. 3. This result extends to a specific weapon type (i.e.  fatalities caused by improvised explosive devices (IEDs)) and to the separate insurgent conflict in Iraq as well as terrorist activity \cite{science} and suggest that the insurgent (and terrorist) production process is very similar in nature across geographical boundaries and borders. It is quite possible that similar results will also be found in the future for cyberattacks.
In the context of a Red-Blue struggle where Red's task of damaging the coalition military effectively resists completion (i.e. Blue is continually fighting back), the observed decrease in completion time may be due to something Red is learning or doing, or  Red's increased manpower, or something Blue is learning or doing (or not doing) or a decrease in Blue's manpower -- indeed there are myriad possibilities. The suggestion of Ref. \cite{clauset} that our progress curve analysis \cite{science} is necessarily tied to insurgent learning or experience, is false.

\begin{figure}
\includegraphics[width=0.95\linewidth]{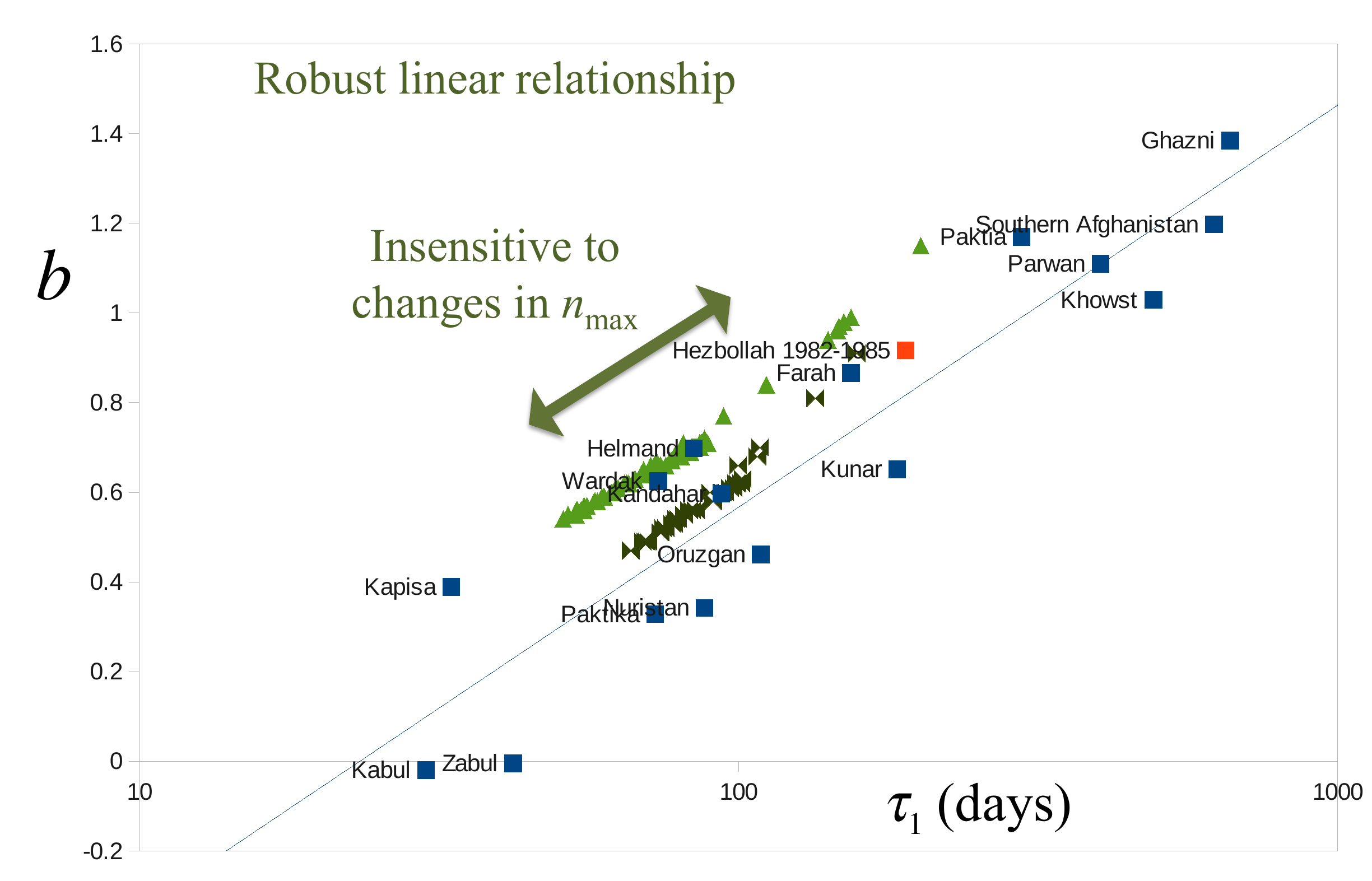}
\caption{(Color online) Robust linear relationship between the escalation parameters for different provinces in Afghanistan shown as blue squares (from Ref. \cite{science}). Also shown (red square)  is a corresponding result for Hezbollah activity (Ref. \cite{science}). In contrast to the claim in Footnote 13 of Ref.  \cite{clauset}, the linear relationship is real, and does not result from the resolution of events on the daily scale: 14 out of the 16 provinces have either zero, one or just two $\tau_n=1$ values. For the two provinces with most $\tau_n=1$ values, Helmand and Kandahar, we demonstrate the robustness by reducing $n_{\rm max}$ (see Fig. 3) such that the amount of piling up of $\tau_n=1$ events is reduced toward zero. There is no qualitative effect on the linear relationship. The green arrow shows that their datapoints undergo a simple to-and-fro variation  {\em along} the line as $n_{\rm max}$ is reduced (Helmand is light green, Kandahar is dark green). For Kandahar, for example, as $n_{\rm max}$ is reduced from the summer 2010 value of 132 down to $n_{\rm max}=24$, for example, the values of $b$ and $\tau_1$ simply  move back and forth in parallel with the line, and at $n_{\rm max}=24$ the datapoint ends up with almost {\em exactly} the same values as for $n_{\rm max}=132$. For Helmand, we show values down to $n_{\rm max}=29$ as an example, as compared to the actual value of 278. For the remaining provinces, removing the very few $\tau_n=1$ values that occur also leaves the linear relationship unchanged. Several provinces have {\em no}  $\tau_n=1$ values, hence the data-resolution criticism is a priori completely irrelevant.}
\label{fig1}
\end{figure}

\begin{figure}
\includegraphics[width=0.95\linewidth]{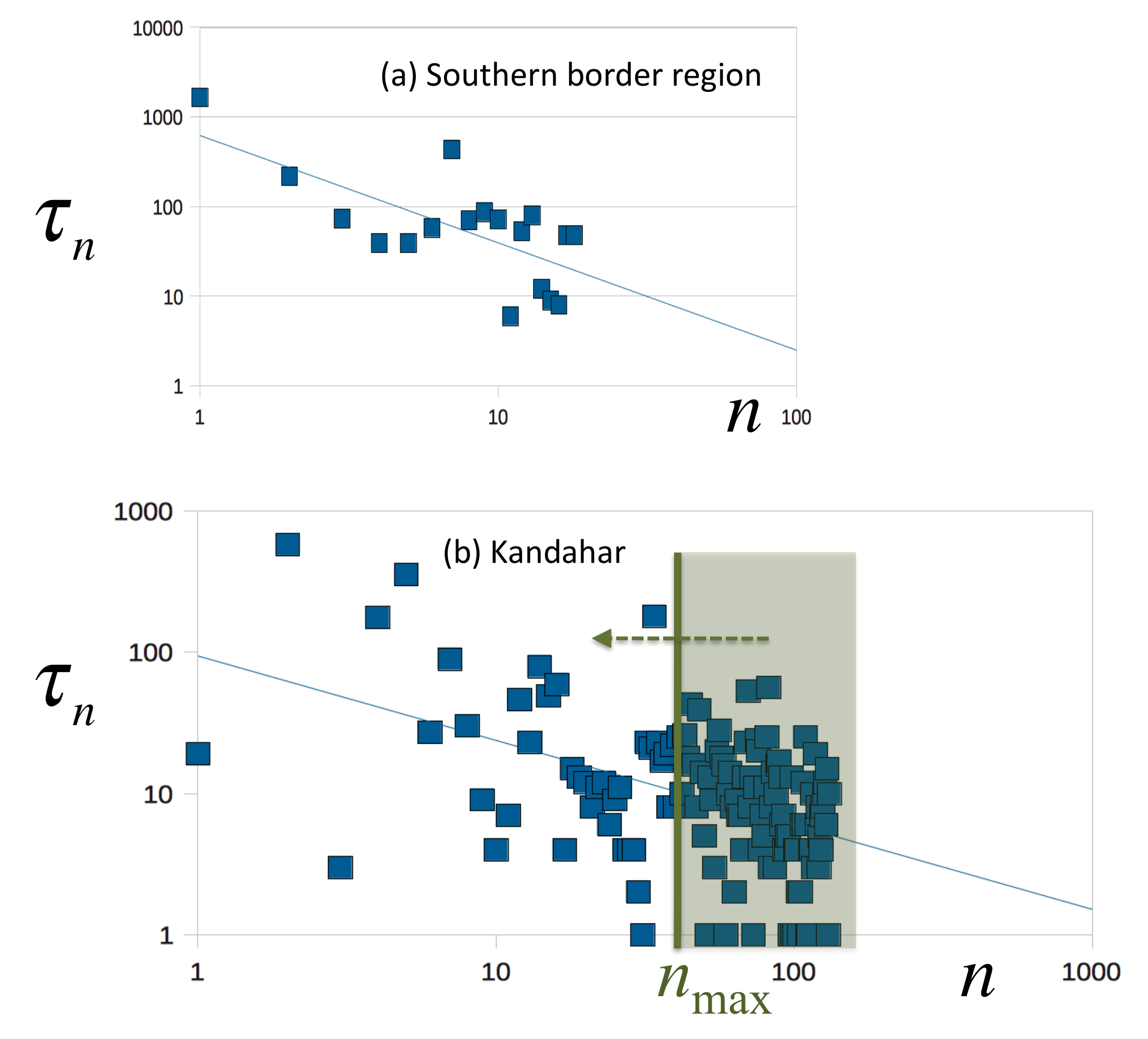}
\caption{(Color online) Successive time intervals $\tau_n$ between fatal days, i.e. days in which coalition 
fatalities are generated by hostile activities, for the Afghanistan province of (a) Southern border region and (b) Kandahar. On this log-log plot, the best-fit power-law
progress curve is by definition a straight (blue) line with slope $-b$ (where $b$ controls the escalation rate). In (a), as in all the provinces other than Helmand and Kandahar, there are either no $\tau_n=1$ values, or just one or two.}
\label{fig1}
\end{figure}

To explain the numbers appearing in Fig. 2, in particular the observed range of $b$ values, we have developed a dynamical version of the Red Queen evolutionary race \cite{science} as shown schematically in Fig. 4. In essence, the theory treats the relative advantage or lead, $R$, of Red over Blue as a stochastic process -- and hence the exponents $b$ are given by the scaling exponent for the standard deviation of the size of $R$'s (possibly correlated) random walk \cite{science}. Using this model, we can interpret the entire spectrum of observed $b$ values for different provinces, and also different terrorism domains, in an intuitive and unified way. Most importantly, this broad-brush Red Queen-Blue King theory does not require knowledge of specific adaptation or counter-adaptation mechanisms, and hence bypasses issues such as changes in insurgent membership (i.e. composition, numbers or numbers of cells), technology, learning or skill-set, as well as removing any need to know the hearts and minds of local residents. Historically, the Red Queen story features the Red Queen (e.g. insurgency or terrorist group) running as fast as she can just to stay at the same place, implying that her Blue King opponent (e.g. state security force or antiterrorism organization) instantaneously and perfectly counter-adapts to her advances such that they are always neck and neck. However, such instantaneous and perfect counter-adaptation is unrealistic -- indeed, the complex adaptation-counteradaptation dynamics generated by sporadic changes in circumstances imply that the temporal evolution is likely be so complex that it can indeed be modelled as a stochastic process. We do not need to know exactly {\em why} $R$ changes at any specific moment, nor do the changes in $R$ have to have the same value since each change is the net result of a mix of factors (e.g. changes in numbers of personnel, technology, learning or experience) for each opponent. We also find that a similar picture to Fig. 2, showing a similarly remarkable inter-relationship between individual provinces, arises in other situations where an arms-race struggle is underway -- for example, for suicide bombings in individual provinces in Pakistan (not shown) \cite{toappear}. In all these cases, we stress that a change in Red's lead $R$ might result from a conscious or unconscious adaptation by Red, or by Blue, or both -- for example, there may be an increase in Red numbers because of a conscious recruitment campaign or simply due to bad press involving Blue's activity. Likewise $R$ may change due to a surge in Blue's numbers or strength, or a change in its tactics or defenses. It does not matter: The precise cause for changes in $R$ does not affect the validity of our theory.

In contrast to the claim of Ref. \cite{clauset}, the linear relationships that we uncover in Ref. \cite{science} and Fig. 2 are real and do not result from any bias due to the resolution of events on the daily scale. Of all the provinces, only Helmand and Kandahar reach an escalation such that events eventually occur on the daily scale. To verify the robustness of our linear relationship, we therefore reduce $n_{\rm max}$ (see Fig. 3) such that the piling up of $\tau_n=1$ events is progressively reduced toward zero. This is equivalent to asking what the progress curve is up to some date prior to our end-date of summer 2010 \cite{science}. As shown in Fig. 2, this has no qualitative effect on the linear relationship -- see the green arrow which indicates the resulting to-and-fro variations of the datapoints for Helmand and Kandahar along the line as $n_{\rm max}$ is reduced. The other provinces have very few, if any, $\tau_n=1$ values. Of the 16 provinces which define the linear relationship in Fig. 1 of Ref. \cite{science}, 14 have two or less $\tau_n=1$ values, and none of these are piled up in a way that would induce a correlation between their best-fit $b$ and $\tau_1$ values. Moreover, as can be checked from media and online reports, the $\tau_n=1$ values that do occur in these 14 provinces can typically be tied to real, separate events on consecutive days, i.e. their one-day separation value ($\tau_n=1$) is real and has nothing to do with data resolution. However for the academic purpose of fully rebutting the claim of Ref. \cite{clauset}, we have gone through the checking exercise of removing these $\tau_n=1$ values in order to test the robustness of the linear relationship between $b$ and $\tau_1$. As the low incidence of $\tau_n=1$ values would suggest, these 14 blue squares follow the linear relationship shown in Fig. 2, and hence also in Ref. \cite{science}, both with and without the inclusion of these one or two $\tau_n=1$ values. Going further, Zabul, Wardak, Khowst and the Southern border region have {\em no} $\tau_n=1$ values and hence this issue is irrelevant to them. Indeed, the fact that these four provinces' best-fit $(b,\tau_1)$ values are widely separated along the linear trendline, means that this linear trend is already well-defined by these four provinces alone. The Hezbollah datapoint (Fig. 2, red) also has no $\tau_n=1$ values and hence there is again no data resolution issue. Even for the organization-wide global terrorist data of Ref. \cite{clauset}, we deduced our best-fit estimates in Fig. 2 of Ref. \cite{science} using the escalation range within which $\tau_n=1$ values were hardly involved (we analyzed the initial escalation in Fig. 2(a) of Ref. \cite{clauset}). In short, the claim in Footnote 13 of Ref. \cite{clauset} that our results are driven by data resolution, is completely false.

We now make a more general point about the recent preprint Ref. \cite{clauset} in relation to our own paper which was published several weeks earlier (Ref. \cite{science}). As stated in its title, our paper focuses exclusively on the escalation regime of fatal attacks against coalition military within individual provinces. The approach in Ref. \cite{clauset}  aggregates over entire organizations, and hence has the potential to produce significant piling-up effects (i.e. densely packed $\tau_n=1$ values) since  the likelihood of a fatal attack happening on a given day somewhere across an entire country or global organization becomes far more likely. Reference \cite{clauset}  treats the resulting regimes of dense $\tau_n=1$ datapoints in an approximate ansatz-driven manner. However this approach \cite{clauset}   has several drawbacks. First, it is actually dangerous to attach too much importance to any regime where $\tau_n=1$ values regularly occur (which we stress is {\em not} the case for our data). A single attack can extend over two days, as has tended to happen in the past decade with the FARC for example in Colombia. However it may erroneously recorded in a terrorism or conflict database as two events separated by $\tau_n=1$. By focusing solely on the escalation period where $\tau_n=1$ events are rare, our paper minimizes this problem, but Ref. \cite{clauset}'s aggregate approach does not.
Second, if events are accelerating, and yet only daily data is known, then eventually the true time interval between fatal attacks will be a matter of hours or minutes, thereby rendering time-series with strings of $\tau_n= 1$ values useless. Imagine a situation where you are monitoring the time between a child's meals -- having a theory (e.g. probability distribution) rounded to an integer number of days is rather pointless since it is a forgone conclusion that each day will involve (at least) one meal.
Third, Ref. \cite{clauset}'s attempt to account for the entire duration using  a single analytic form, runs the risk of sacrificing improved accuracy and insight during the practically important regime of initial escalation (i.e. small $n$). By focusing on this escalation regime, we were able to use a far simpler analytic form than Ref. \cite{clauset} and hence were able to discover the remarkable linear dependencies shown in Fig. 2 and Ref. \cite{science}, as well as developing a potentially powerful Red Queen theory which is not tied to any specific mechanism (e.g. insurgent size). Instead, Ref. \cite{clauset} forces the establishment of a single analytic form to cover both the escalation regime with $\tau_n\gg 1$  and the steady-state-like regime with $\tau_n\approx 1$. This is problematic, since a new conflict in its escalation phase does not `know' that it is approaching the data-resolution limit $\tau_n= 1$, hence there is no physical reason that its mathematical description should involve a smooth trajectory through the $\tau_n= 1$ resolution boundary.
We also note that the occasional jumps that may be observed in the time interval between successive fatal days (see for example, Fig. 3) might be interpreted as systematic disruptions by Red or Blue on successive days (e.g. breaking the daily routine) as opposed to being measurement error or background fluctuations. Such jumps may arise for any number of reasons including (but not limited to) changes in the number of insurgents or cells, and may even act like memory resets to the  process. It is these jumps that we believe could yield valuable insight into the effects of  hidden changes in the operating landscape on both sides of a conflict, either in real or cyber space. We will return to this in future work.

\begin{figure}
\includegraphics[width=0.95\linewidth]{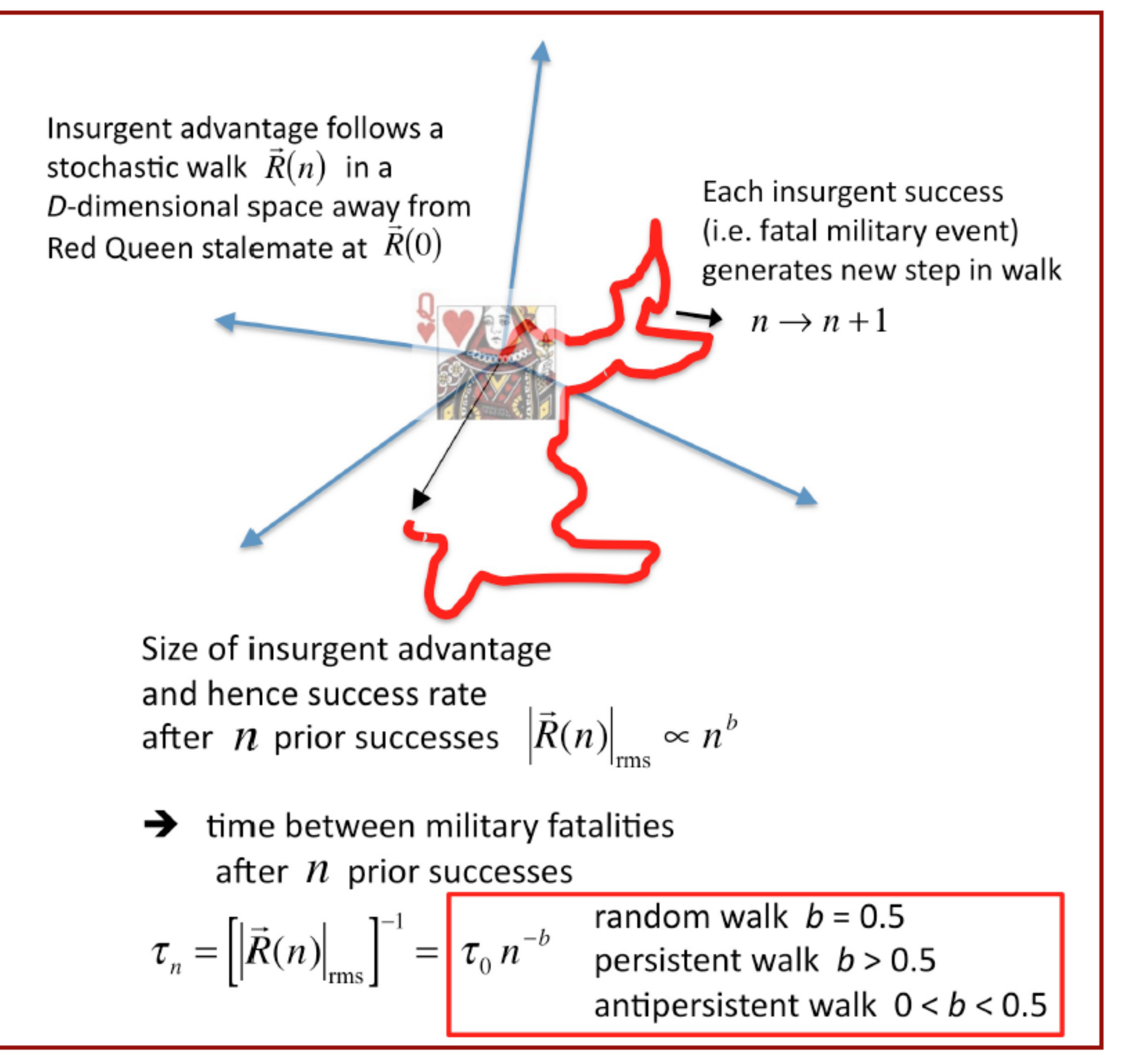}
\caption{(Color online) From Ref. \cite{science}, our dynamic Red Queen model for the Red-Blue struggle. Red (e.g. insurgent) advantage
$R$ is represented as a vector in a multi-dimensional space whose axes may represent technological, psychological, social, cultural or behavioral factors. $R$ follows a stochastic walk in this $D$-dimensional space. Using known results from statistical physics, exact results can be obtained for $b$
under different conditions of correlation etc. within the walk. For the simplest case of an uncorrelated walk, $b=0.5$. For a completely correlated walk in $D=1$ dimensions (i.e. linear feedback), $b=1$. For a mean-reverting walk in $D=1$ dimensions, $b\approx 0$.}
\label{fig1}
\end{figure}

Reference \cite{clauset} presents additional false criticisms of Ref. \cite{science}. For example, it claims that we use learning to explain the progress-curve escalations. This is not true -- there is no such statement in Ref.  \cite{science}. On the contrary, we state explicitly in Ref. \cite{science}  that `Our broad-brush theory does not require
knowledge of specific adaptation or counteradaptation
mechanisms, and hence bypasses issues
such as changes in insurgent membership,
technology, learning, or skill set, as well as a need
to know the hearts and minds of local residents.' 
We also state in Ref. \cite{science} that `We do not need to
know exactly why $R$ changes at any specific
moment, nor do the changes in $R$ have to have
the same value, because each change is the net
result of a mix of factors'. Indeed, we purposely chose the term `progress curve' to avoid any explicit connection to learning. As we acknowledge in Ref. \cite{science}, progress can arise for many different reasons (including, but not limited to, increases in insurgent numbers). Neither our Red Queen theory nor our results depend on learning, nor do they rely on restricting the scope of possible driving mechanisms. 
Insurgent size increase, as promoted in Ref. \cite{clauset}, is simply one of the many specific mechanisms that can increase the Red Queen's lead $R$ and hence escalate fatal attacks. As Ref. \cite{science} states, the power of our Red Queen theory as opposed to a specific, yet unjustified, mechanism such as insurgent size increase, is that it is not tied to one particular narrative. It properly allows for combinations of mechanisms which may change over time. Indeed, the claim in Ref. \cite{clauset} that organizational growth drives the escalation, is highly problematic in terms of verification. No data exists -- nor will any ever likely exist  -- for reliable insurgent numbers within the individual provinces throughout the entire duration of our study. A similar story holds for other insurgent wars and terrorism \cite{clauset}. In particular, estimates of entire organizational size (e.g. the total number of Taliban) are extremely crude at best, and may in fact be misleading -- in particular, it is unclear whether they bear any relationship to the actual number of active members who are ready to carry out attacks at a particular moment in time.

The explicit explanation of escalation suggested by Ref. \cite{clauset}, invokes a change in the size of the organization. However, earlier in 2009, Ref. \cite{nature}  had already introduced a detailed model (summarized below and in Figs. 7 and 8) in which the number of active cells $N_g$ (i.e. semi-autonomous groups \cite{nature}) can increase or decrease over time. We had then used this to explain the change in the daily frequency distribution of fatal attacks on civilians etc. aggregated at the level of a country (see Fig. 5). This model \cite{nature} provides a quantitative explanation of the temporal evolution of several conflicts in terms of an increase (decrease) in the total number of insurgent groups (i.e. cells) over time {\em and} an effective lowering of the bar for carrying out successful attacks. This finding \cite{nature} reinforces our argument above against size driving escalation, in that it is not enough to just focus on the changing number of cells (or total number of insurgents). For a detailed discussion, see the online Supplementary Information of Ref. \cite{nature}, with results in Supplementary Table 2, and explicit model flow-chart in Supplementary Figure 5. Even the basic one-population version of our model (see Appendix A) can have the number of agents increasing, or more generally changing, over time without affecting the appearance of an approximate power-law distribution with slope near 2.5 (Fig. 6). 
Going further, the following simple argument reinforces our claim that organizational growth cannot be the sole driver of the escalation patterns that we observe in Fig. 3. Suppose at time $t$,  insurgents (or terrorists) have a strength $N_R(t)$ while coalition troops (or a counterterrorism force) have strength $N_B(t)$, where `strength' might involve many factors, but for simplicity we take it to be the number of agents on each side. When $N_B(t)\approx 0$ (which can arise temporarily in some provinces as troops get shifted around) the rate of fatal attacks on coalition military will, by definition, be essentially zero, irrespective of the size of $N_R(t)$. Even if $N_B(t)\gg 1$, there will also be essentially zero fatal attacks if these coalition military never go out on patrol. Hence a blanket statement that insurgent size dictates escalation is wrong. Suppose instead one tries setting the rate of fatal attacks -- taken to be proportional to the Red Queen's lead -- as $R(t)=k (N_R(t) - N_B(t))$ where $k$ is some conflict-specific constant. One can trivially see that increasing the size $N_R$ does not guarantee an increase in $R$ and hence attack frequency, because $N_B$ may also change. Similar conclusions hold for other functional forms, since the generation of Blue fatalities requires by necessity some kind of $N_B(t)$ dependence. Hence even before extra complications such as changes in tactics or equipment on either side etc. are added, one can see that a theory based solely on $N_R(t)$ is not plausible.

\begin{figure}
\includegraphics[width=0.95\linewidth]{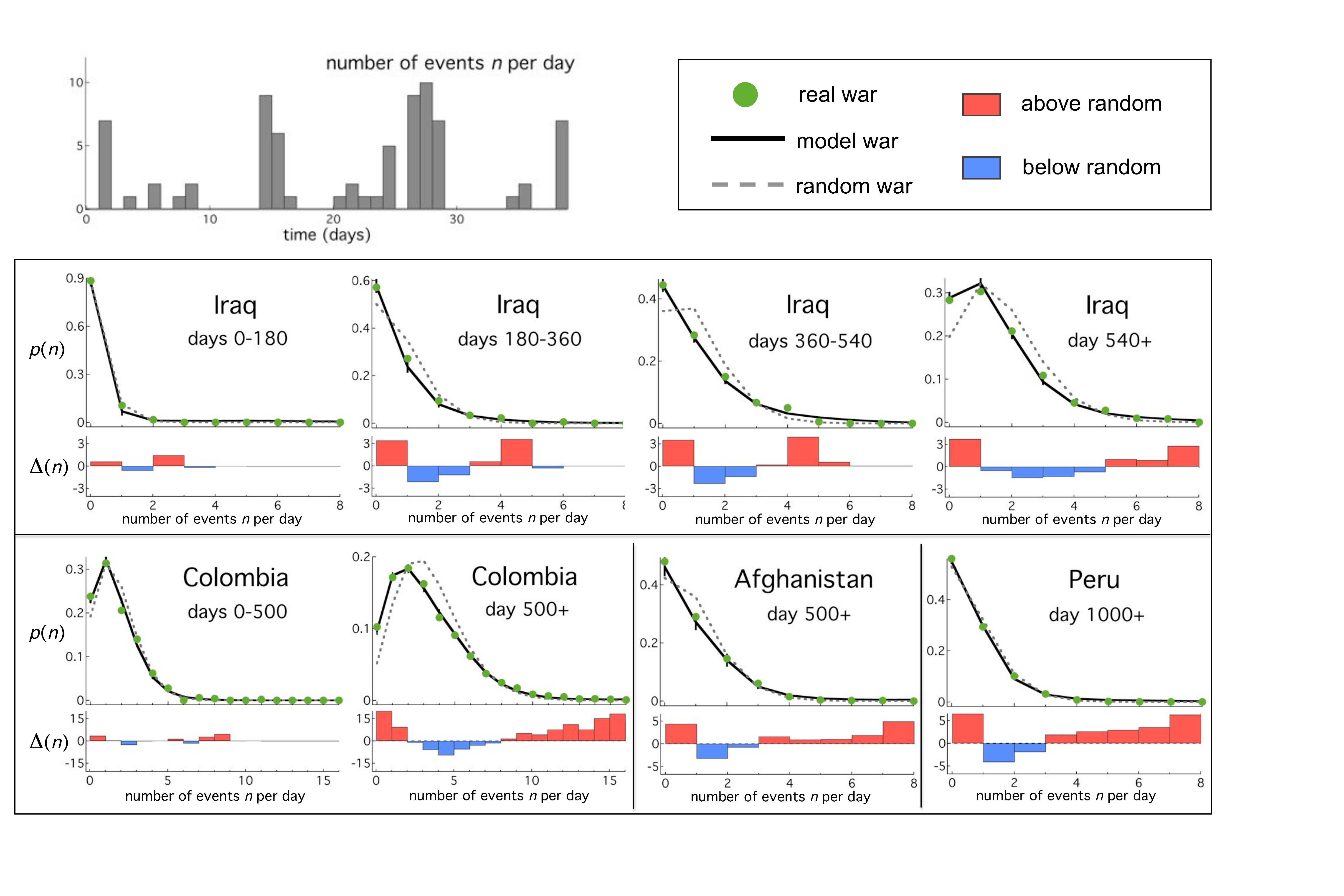}
\caption{(Color online) From Ref. \cite{nature}, the distribution of the number of violent events per day in a given conflict. Green circles show the distribution $p(n)$ for the number of days with $n$ events in the actual conflict. Histograms below represent differences $D(n)$ between real and random wars, in units of standard deviations from the mean. The average values for random wars (i.e. where actual data is randomized over finite time window) are shown as dashed lines. Solid lines denote average distributions calculated from 10,000 realizations of our model. This model is a generalized version of the {\em El Farol} model: Individual cells only act if they possess strategies with sufficient past success, and hence the cell surpasses a minumum confidence level \cite{nature,book}. Error bars
for random wars, calculated as one standard deviation from the mean of 10,000
shufflings, are shown but they are small. The error bars for the model wars demonstrate a small spread in run outcomes.}
\label{fig1}
\end{figure}

\begin{figure}
\includegraphics[width=0.95\linewidth]{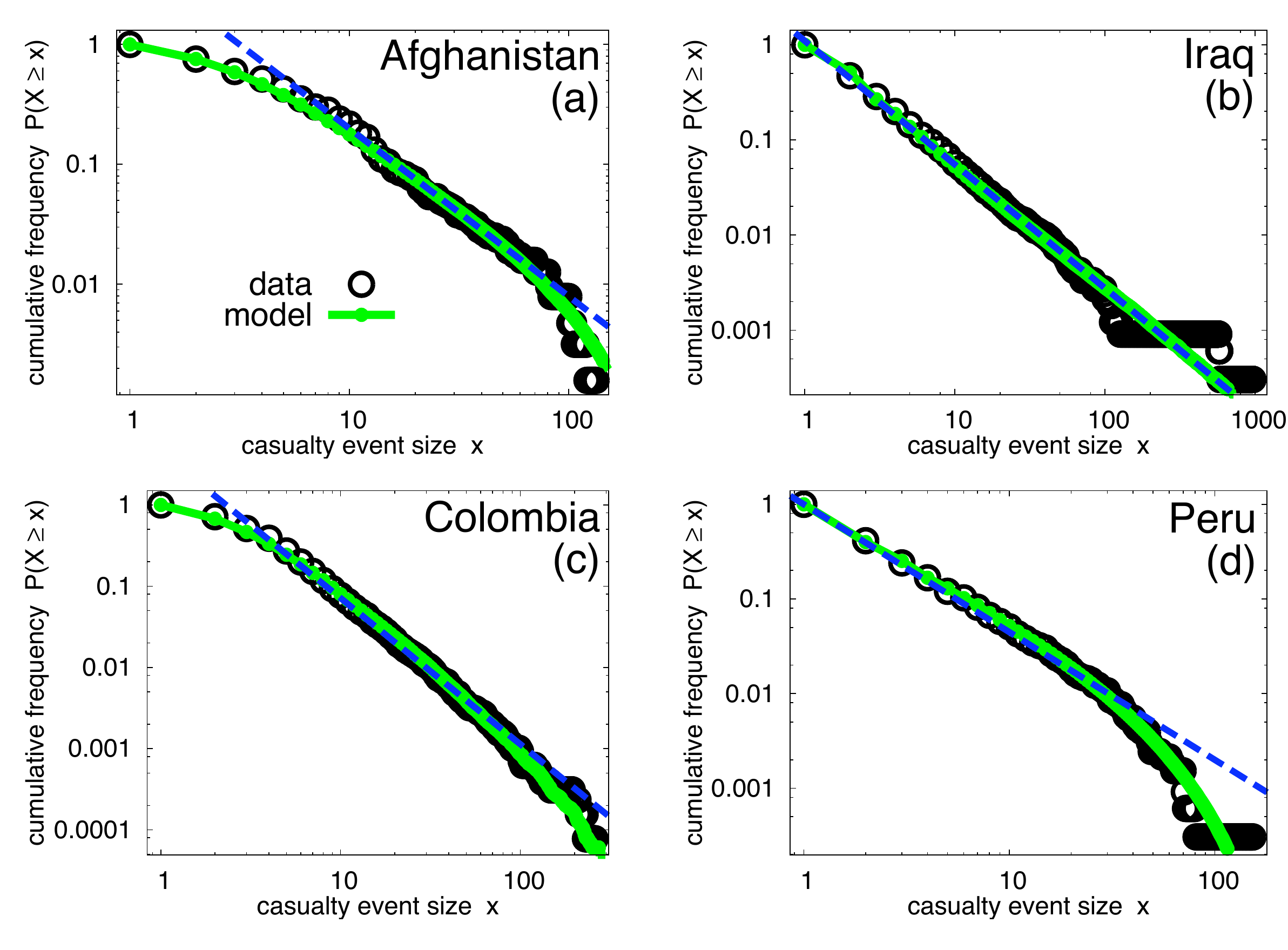}
\caption{(Color online) From Ref. \cite{nature}, the log--log plot of the complementary
cumulative distribution of event size $P(X\geq x)$ (i.e. the probability of an
event of size greater than or equal to x) for four conflicts. Horizontal axis shows event size x, namely the number of casualties. Solid
green curves show the results from our model. Blue dashed line is a straight
line guide to the eye, not a power-law fit. As also shown in Ref. \cite{nature}, the hypothesis that these quasi-straight-line graphs follow a power-law cannot be rejected. The exponents for the underlying best-fit power-laws, all have values near $2.5$. This can be explained using a simple, one-population version of our model (see Appendix A), assuming that a given insurgent cell inflicts damage proportional to its size when it attacks. However, to fully understand the richness of the full distribution, and in particular to understand the deviations from power-law, the activity of the other actors (i.e. state military, police or civilians) must be accounted for. Doing this yields remarkably good agreement with the empirical data (green curves) \cite{nature}.}
\label{fig1}
\end{figure}

We also flag the incorrect discussion in Ref. \cite{clauset} concerning our model's ability to account for the lack of a strong empirical relationship between the severity of individual attacks and the escalation in the number of attacks \cite{clauset}.  Our full model (Fig. 8) does not in fact contradict this finding. We had already shown that the best-fit power-law exponent does not change much over successive epochs \cite{lanl1,lanl2,aps,wars} even though the frequency distribution for the number of events per day is changing (Fig. 5) \cite{nature}.
Turning to the basic one-population version, the severity of an attack is given by the size of the cell deciding to perform the attack, while the size of the entire insurgent or terrorist organization is given by $N(t)$. Figure 7 shows an example where prior to fragmentation of the cell of size 3 into 3 cells of size 1, $N(t)$ is partitioned in such a way that $n_{s=1}(t)=0$, $n_{s=2}(t)=1$, $n_{s=3}(t)=2$, $n_{s=4}(t)=0$, $n_{s=5}(t)=1$, $n_{s\geq 6}(t)=0$. The total size is $N(t)=\sum_s n_s(t)=1\times 2+2\times 3+1\times 5=13$, and the number of cells $N_g(t)=4$. After fragmentation of the cell of size 3 into 3 cells of size 1, $N(t)=13$ still, but now $N_g(t)=6$ and we have $n_{s=1}(t)=3$, $n_{s=2}(t)=1$, $n_{s=3}(t)=1$, $n_{s=4}(t)=0$, $n_{s=5}(t)=1$, $n_{s\geq 6}(t)=0$. Hence the actual values of $N(t)$, $N_g(t)$ and $n_s(t)$ (the number of cells with size $s$) can show appreciable variability from each other at any particular moment in time. In passing, we note the interesting feature (see Appendix A) that the actual functional form of the cell size distribution of individual cells $\{n_s(t)\}$, where $n_s(t)$ is the number of cells of strength $s$ at time $t$,  does not depend on $N(t)$. 

Finally, we mention for clarification that our escalation study in Ref. \cite{science} looked at the deterministic (power-law) trend of the sequence of time intervals. We did not look at a statistical distribution of time intervals, nor did we analyze power law distributions -- nor does Ref. \cite{science} claim that the time intervals are described by a power-law distribution. There is a fundamental difference between describing a deterministic overall trend in successive time intervals as the conflict evolves  -- which is what we did for each province -- and describing the statistical distribution of the ensemble of time intervals for each province. Since the conflict is non-stationary, and shows escalation, it would make little sense to aggregate all the time intervals and look for a single distribution. This would ignore completely the underlying dynamical trend, and could lead to quite misleading conclusions. For example, the few large $\tau_n$ values during the early stages of the conflict would tend to dominate any statistical analysis of fat-tailed behavior, giving a distribution measure which is unrepresentative of the entire conflict. On another point related to statistics, and specifically returning to the conventional unweighted linear least-squares approach which we used in Ref. \cite{science} to detect the trend in ${\rm log} {\tau_n}$ versus ${\rm log} n$, it is well-known that this method will become very accurate in the limit that the residuals of ${\rm log} {\tau_n}$ approach statistical independence with identical distributions (i.i.d.). This i.i.d. criterion is not strictly met in many applications of least-squares in the sciences, however it turns out to be a good approximation in our study \cite{science} and likewise in Figs. 2 and 3. This is because the error in the underlying $\tau_n$ values has a crudely multiplicative form whose effect decreases with increasing $n$. There are many reasons why such a scatter should occur: Insurgent attacks early in the conflict may stop short of wanting to cause coalition military deaths for fear of stimulating an increased future troop presence, thereby producing large variations in $\tau_n$ values for small $n$. Also, there were physically less coalition soldiers (i.e. targets) on the ground, so their fatalities may have been more clustered. Because of the subsequent transformation to logarithmic variables, the resulting residuals for ${\rm log} {\tau_n}$ versus ${\rm log} {n}$ will then have distributions which are reasonably insensitive to increasing $n$. It also turns out that these residuals have a fairly small autocorrelation, a fact that can also be seen crudely by eye simply by looking at the scatter of ${\rm log} {\tau_n}$ values. The net result is that the residuals of ${\rm log} {\tau_n}$ exhibit a distribution which is fairly insensitive to $n$ and they also have very little autocorrelation -- in short, the residuals are approximately i.i.d. as indeed they should be. 

\begin{figure}
\includegraphics[width=0.85\linewidth]{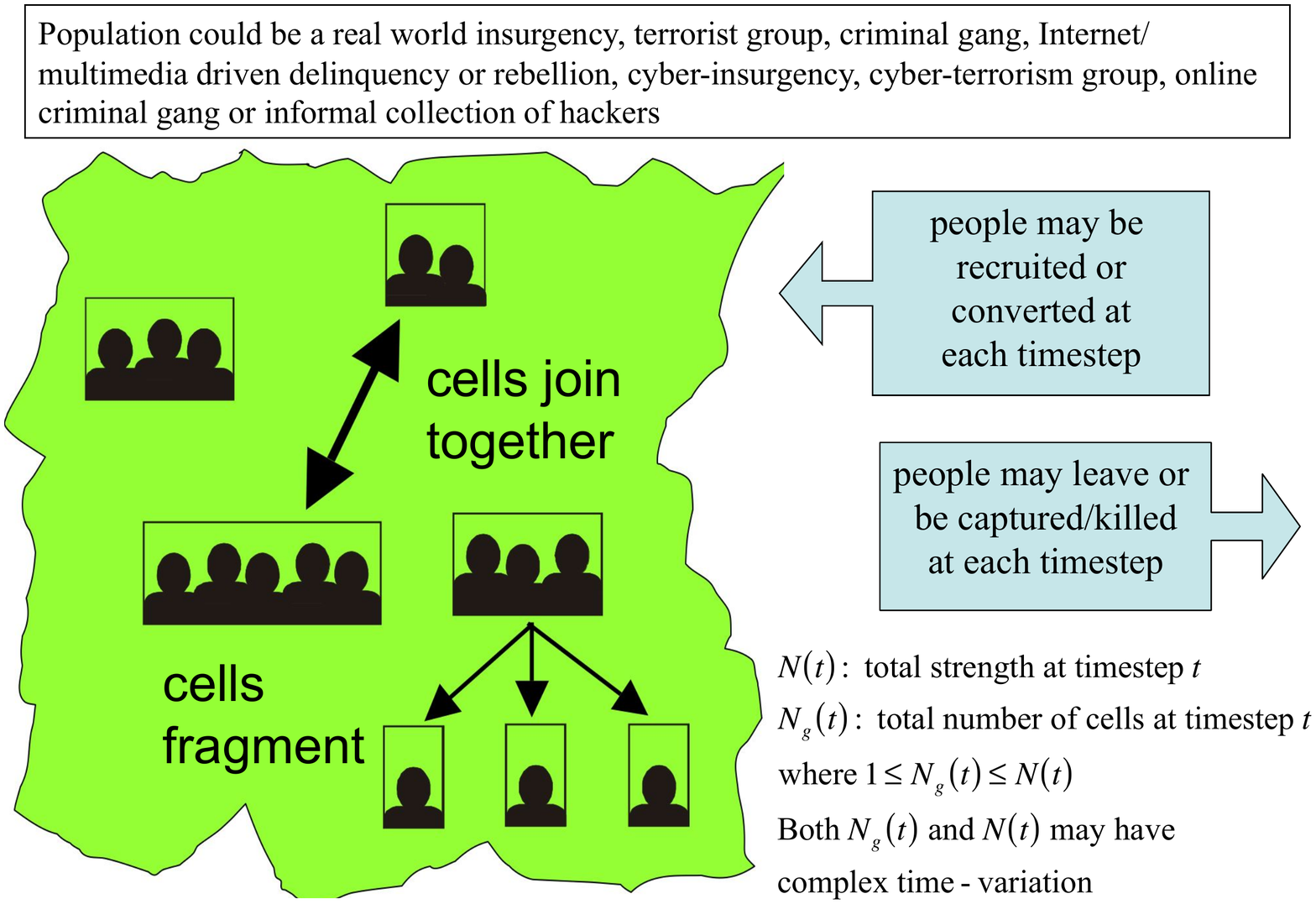}
\caption{(Color online) From Ref. \cite{nature}, the insurgent population
comprises an overall strength $N(t)$, distributed into dynamically evolving cells -- with time-varying size, number and composition, and with diverse
strengths at each time-step $t$. The total number of cells $N_g(t)$ at time $t$ can vary with time, as can the total number of composite objects (i.e. insurgent members, equipment, information) $N(t)$.  Since $N_g(t)$ is the number of cells, and $N(t)$ is the total number of objects (e.g. insurgents) these two quantities are fairly independent with the only constraint being that $N_g(t)\geq 1$ (i.e. the smallest number of cells is when every object belongs to this same cell) and $N_g(t)\leq N(t)$ (i.e. the largest number of cells is when every object is isolated). In this example shown, the number of cells of a given size $s$ at this timestep $t$, prior to fragmentation of the cell of size 3 into 3 cells of size 1, is $n_{s=1}(t)=0$, $n_{s=2}(t)=1$, $n_{s=3}(t)=2$, $n_{s=4}(t)=0$, $n_{s=5}(t)=1$, $n_{s\geq 6}(t)=0$. The total number of insurgents is $N(t)=\sum_s n_s(t)=1\times 2+2\times 3+1\times 5=13$. The number of cells $N_g(t)=4$. After fragmentation, $N(t)=13$ still, but now $N_g(t)=6$.}
\label{fig1}
\end{figure}

\begin{figure}
\includegraphics[width=0.7\linewidth]{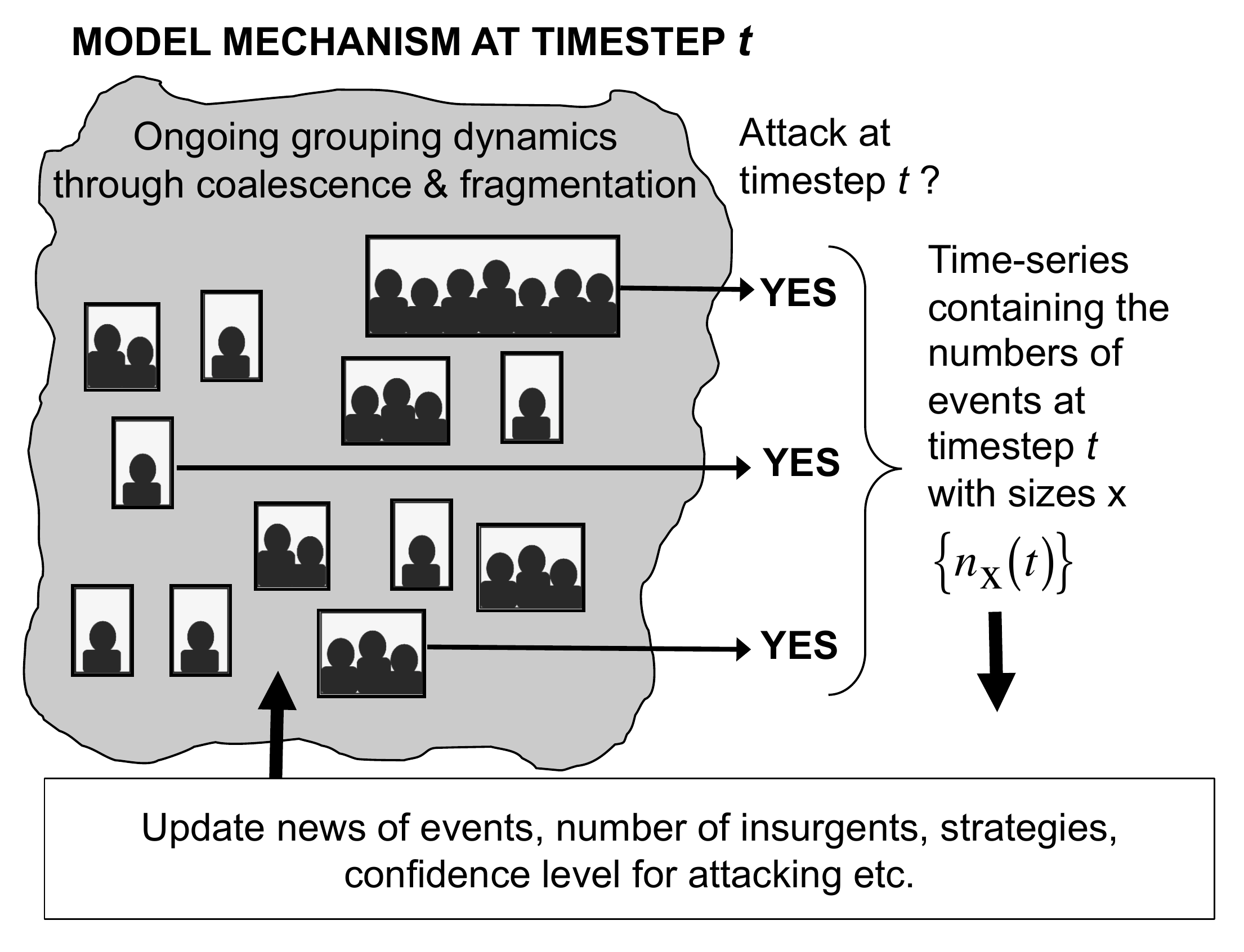}
\caption{(Color online) From Ref. \cite{nature}, our full model describing the attack severity distribution and attack timings comprises two main processes: (1) The interaction between Red (e.g. insurgent force, shown here in black) of size $N(t)$ and Blue (e.g. coalition military, not shown) and Green (e.g. civilian population, not shown) which then generates the severity distribution for casualties per event. As a result of these interactions, the insurgent force of size $N(t)$  undergoes an ongoing process of coalescence and fragmentation. (2) A repeated decision process in which each of the individual $N_g(t)$ cells which exist at time $t$, assess whether to attack or not based on the information and resources it has available to it. This is a two-step process: First the cell will either be potentially active or not \cite{nature} based on its confidence level (e.g. based on the performance of of their strategies in the past). Then if active, it will decide whether to specifically attack at that timestep or hold off momentarily. The fact that the severity of events and their timing show a low correlation in our datasets, is reflected in the fact that $N_g(t)$ and $N(t)$ can vary almost independently, with the only constraint being that $N_g(t)\geq 1$ (i.e. the smallest number of cells is when every object belongs to this same cell) and $N_g(t)\leq N(t)$ (i.e. the smallest number of cells is when every object belongs to this same cell). In this example, the number of cells of a given size  $s$ at this timestep $t$ is $n_{s=1}(t)=6$, $n_{s=2}(t)=2$, $n_{s=3}(t)=3$, $n_{s=4-6}(t)=0$, $n_{s=7}(t)=1$, $n_{s\geq 8}(t)=0$. The total number of insurgents is therefore $N(t)=26$. The number of cells $N_g(t)=12$.}
\label{fig1}
\end{figure}

\section{ATTACK TIMING: NON-POISSON DISTRIBUTION}
Figure 5 reproduces a result from Ref. \cite{nature}, in which we showed that the empirical data for the timing of attacks generates non-Poissonian distributions. In line with the above discussion, the non-Poisson nature of the distribution for the number of attacks on a given day of an ongoing conflict, tends to increase as the conflict evolves over time. Figure 5 also demonstrates a common burstiness in the distribution for the number of attacks per day. As explained in the
Methods and Supplementary Information of Ref. \cite{nature}, we compare the distributions
over daily event counts for different epochs within four modern
conflicts, against control distributions (Ôrandom warÕ) obtained by randomizing event occurrences within each epoch. The data for each conflict (green circles) deviate from its random war (dashed curve) in a similar way: the real war exhibits an overabundance
of light days (i.e. days with few attacks) and of heavy
days (i.e. days with many attacks), but a lack of medium days
compared with the random war (see lower panel of Fig. 5). By considering
subsets of days, we determined that these features are not just an
artefact of a variation in attack volume across days of the week (for
example, Fridays). The burstiness became more pronounced over time for the wars in both Iraq and Colombia, suggesting that they became less random as
they evolved. We checked that these findings are insensitive to the precise specification
of the epochs within a given conflict.

\section{SEVERITY: POWER-LAWS AND BEYOND}

As mentioned earlier, Refs. \cite{lanl1,lanl2} showed that the distributions for the severity of attacks (i.e. the histogram of the number of events with $x$ casualties, as a function of $x$) generates surprisingly broad and common distributions. These distributions approximate to a power law -- or strictly speaking, a power-law cannot be rejected -- and have a corresponding power-law exponent around 2.5. However in Ref. \cite{nature}, we looked at features beyond a simple power-law and found that additional information is contained in the deviations beyond the strict power-law form. In particular, we generated successively more detailed models in which populations of actors interacted over time, like an ecology, and the output of these interactions gave the casualty distributions. The good fit between these models and the empirical deviations beyond power-law (see Fig. 6), offers new insight into  subtle differences in the rules-of-engagement for these conflicts.

\section{OUR MICROSCOPIC MODEL OF MULTI-ACTOR CONFLICT}
There is much work across the disciplines on how groups interact, and how these groups break and form in non-violent settings -- for example, in social psychology for humans, and evolutionary biology and zoology for animals, birds and fish. There is far less known about how human groups' joining and breaking processes change when they are operating in a clandestine and/or illicit way such that they do not want to get caught, and also where they may have an underlying mistrust of each other which changes over time \cite{gambetta,robb,kenney}. However, clues can be gathered from a number of related areas, including animal anti-predator behaviors \cite{caro,ian2} and also from the study of criminal gangs \cite{felson,gambetta,kenney}. All these works -- in addition to journalistic reports for situations as diverse as violence in Colombia, Iraq and Afghanistan, through to the recent riots in London -- point to the idea that there are many actors involved, and they form loose, ethereal groups which are hard to detect and whose internal structure is either continually in flux or changes sporadically. 
These features inspire the relatively few, yet entirely reasonable, mechanisms that we adopt in Fig. 8 in order to define a mechanistic model: namely, coalescence and fragmentation for the cell dynamics, and bounded rationality in the decision-making for the cell attack decisions. Indeed, the fact that there is only weak correlation between the severity of events and their timing, further supports the idea that the severity and timing mechanisms can be crudely thought of as independent processes, to a zeroth order approximation. 

In Ref. \cite{nature}, we presented a full model which reproduces the entire severity distribution across various insurgent conflicts (green curves in Fig. 6). In this full model, Red and Blue interact and fatalities are inflicted in Red, Blue and Green. Our results show that to fully understand the richness of the severity distribution, and in particular to understand the deviations from power-law, the role of Red, Blue and Green must be accounted for. Doing so yields remarkably good agreement with the empirical data over the entire distribution (green curves in Fig. 6). However if one is simply looking for an explanation of the approximate power-law behavior of the distribution's tail, and the apparent ubiquity of power-law exponents for casualties in the region of 2.5 for both insurgencies and terrorism, we also showed \cite{nature} that a simple, one-population version of our model will suffice (see Fig. 7). Appendix A gives an explicit derivation for the resulting distribution of Red cell sizes $\{n_s(t)\}$. Assuming that a given insurgent cell inflicts damage proportional to its size when it attacks, the empirical observation is reproduced concerning the approximate power-law for casualties with exponent near 2.5.

Our full model \cite{nature} (Fig. 8) combines two key human behavioral features: 

\noindent (1)	{\em Bounded rationality in the decision-making process} of a cell when it is deciding whether to attack on a given day. As in the famous `El Farol' bar problem \cite{book,wang}, the cells are  limited by the information they have available to them, and the time they have to make their decisions. In the `El Farol' bar problem \cite{book,wang}, a collection of boundedly rational agents are each deciding whether to attend a potentially overcrowded bar, and hence are deciding whether today is a good day to compete for the limited seating space. In the case of the insurgent cells, they are each deciding whether today is a good day to attack -- the limited resource that they are competing for could be space in the international news headlines. This interpretation is consistent with the behavioral feature of insurgents noted by former U.S. Senior Counterinsurgency Adviser David Kilcullen \cite{packer} that when insurgents ambush an American convoy in Iraq: ``...they're not doing that because they want to reduce the number of Humvees we have in Iraq by one. They're doing it because they want spectacular media footage of a burning Humvee." It is also consistent with small-scale laboratory experiments studying human groups  \cite{wang}. Further support for this feature of boundedly rational decision-making, is provided by Kenney \cite{kenney}: ``Éparticipants in trafficking networks are only what Herbert Simon calls boundedly rational: they face significant computational limitations in their ability to analyze feedback from incoming stimuli". 

\noindent (2)	{\em Fragile dynamical clustering} within an insurgent population (e.g. as a result of internal interactions and/or the presence of an opposing entity such as a state army), just as schools of fish or animals will go through cycles of build up and then rapid dispersal when a predator approaches \cite{caro,ian2,gambetta,robb,kenney}. The coalescence-fragmentation process (see Figs. 6 and 7) is consistent with current notions of modern insurgencies as fragmented, transient, and evolving \cite{bbcColombia,kilcullen,robb,packer,palla}. We recall the phrase of Gambetta \cite{gambetta} ``.... contrary to widespread belief, criminal groups are unstable." Further support is again provided by Kenney \cite{kenney} in {\em From Pablo to Osama: Trafficking and Terrorist Networks, Government Bureaucracies, and Competitive Adaptation}: ``To protect themselves from the  police, trafficking enterprises often compartment their participants into loosely coupled networks and limit communication between nodes''; ``Trafficking networks . . . . are light on their feet. They are smaller and organizationally flatter"; ``In progressive-era New York, according to historian Alan Block, cocaine trafficking was organized by different networks of criminal entrepeneurs who formed, reformed, split, and came together again as opportunity arose and when they were able"; ``loose collection of ÔcellsÕ containing relatively small number of cell workers"; ``Abu Sayyaf . . operates as a decentralized network of loosely coupled groups that conduct bombings, kidnappings, assassinations, and other acts of political violence in pursuit of a common goal . . ". Kenney also highlights the close connection of traffickers to terrorists: ``Al Qaeda share numerous similarities with drug-trafficking enterprises" \cite{kenney}.

The coalescence process mimics the situation in which two
cells (or individuals in these cells) initiate a communications link between them of arbitrary
range (for example, a mobile phone call), and hence the two cells tend to coordinate their actions from then on -- albeit maybe loosely. Indeed, the individual agents  need not know each other, or be physically present in the same place. The long-range nature of the coupling makes it a reasonable description for physical insurgencies and crime groups using modern communications in real space, as well as cells acting in cyberspace -- or any mix of the two \cite{kenney}. Indeed, the language of what is a cell and what is a group, and what is crime and what is insurgency, becomes somewhat irrelevant since the mechanistic operational details are now very similar. Further details are given in Ref. \cite{nature}.
The fragmentation process (Fig. 6) may arise for a number of social or situational reasons, from breakdown in trust within the cell \cite{gambetta} through to detection of imminent danger \cite{caro,kenney}. 
In addition to the quotes above concerning insurgent, drug and criminal groups \cite{kenney}, it is well documented that groups of objects (e.g. animals, people) may suddenly scatter in all directions (i.e. complete fragmentation) when its members sense danger, simply out of fear \cite{caro} or in order to confuse a predator \cite{caro}. Or they may fragment following a clash in which the cell perceives that it is losing (see Supplementary Information in Ref. \cite{nature} and Refs. \cite{wars,blazej} for a number of variants, all of which give similar empirical distributions for the severity). 
Interactions are distance-independent as in Ref. \cite{ez} since we are interested in systems where messages can be transmitted over arbitrary distances (e.g. modern human communications). Bird calls and chimpanzee interactions in complex tree canope structures can also mimic this setup, as may the increasingly longer-range awareness that arises in larger animal, fish, bird and insect groups \cite{caro}. 
Appendix A gives an illustration of the type of mathematical analysis which is possible, for the basic version of our model, stripped down to a simple form with no decision-making, and only one population -- the Red insurgency. Instead of having cells fragment when interacting with Blue, or when sensing imminent danger, we simply assign a probability for them to fragment. The resulting model yields an exponentially cutoff 2.5-exponent power-law for the distribution of group sizes. Assuming that the civilian population is just some passive background that receives an impact proportional to the strength of each cell when it acts, the distribution of civilian casualties will also have this same distribution -- which is indeed what is observed for a wide range of insurgent conflicts and global terrorism, as reported in Ref. \cite{nature}. 

Reference \cite{clauset2009} makes a  claim that our coalescence-fragmentation model falls down on the basis that an approximate power-law severity distribution exists from the outset of their empirical dataset for each terrorist organization, and yet the coalescence-fragmentation process surely needs some time to converge to its steady-state power-law distribution. However, this claim is false. First, the $N(t)$ initial members may be coalescing and fragmenting before any violent event is undertaken -- indeed, there are many examples of underground organizations and US-based militia who spend many years evolving without any noticeable violent activity. No external event may be observed, but there is still a dynamical network of groups evolving in the background. Most importantly, any such organization will undoubtedly already have several existing clusters of contacts, hence it is not the case that the distribution has to build up from all isolated agents. Going further, it is well known that an approximate power-law for group sizes with slope around $2.5$, as produced by our model, is to be expected with many different social and human activity scenarios -- from the way people organize themselves in markets \cite{Gabaix}, to commerce \cite{felix}, through to more casual social settings. A nascent insurgent, criminal, or cyber group could be created effectively instantly from such an existing structure. 
Second, the numerical simulations show that the fat-tailed distribution develops very quickly, even if we start with isolated agents. 
Third, it is not the case that starting from day 1 of a given organization, all fatal events are recorded in the database. There is no guarantee that the finite time-window database of Ref. \cite{clauset}, which started nominally in 1968, either records correctly all events since 1968, or that it captures the true first few events of each terrorist group. The way in which events are interpreted and recorded has changed over time, and so in addition has the ability to name organizations -- and indeed, so have their names changed in some cases, with merging and splinter group formation fairly common. We also note that the alternative severity model  proposed in Ref. \cite{clauset1}, is simply a combination of phenomenological broad-brush factors which happen to give a power-law, but without any specific justification for yielding the observed exponent value of 2.5. Instead, the parameters of this model \cite{clauset1} need to be cherry-picked in order to obtain the observed power-law exponent value of 2.5. In reality, a continuum of values -- including values well away from 2.5 -- are just as likely within the model \cite{clauset1}. Nor is there any quantitative evidence to support their proposed underlying mechanism.

We note that we have also carried out preliminary investigations (and are now pursuing rigorously) the addition of heterogeneity in terms of individual ÔcharacterÕ, as in Fig. 1, and its effects on team formation and kinship when both the individuals and cell are under stress. We have already presented this work for gangs and online guilds for massively parallel human activities involving online cyberwar games \cite{gangs}, as well as investigating the effect theoretically in a preliminary way \cite{blazej}. This published work successfully uses the addition of a scalar character variable to describe the empirical datasets for Long Beach street gangs and World of Warcraft online guilds. Enriching this structure, our preliminary work suggests that the inclusion of agent character may cause mixing of these divisions and initiate extreme behavior, depending on the strength of the kinship tendency (e.g. mimicking tribal and ethnic tendencies). Eventually, we hope that a full character-version of our model will provide a flexible tool which we can adapt to help address a number of issues concerning social and cultural intervention schemes, such as ceasefires and peace plans, and pinpointing social triggers that aggravate a given conflict. 

\begin{figure}
\includegraphics[width=0.95\linewidth]{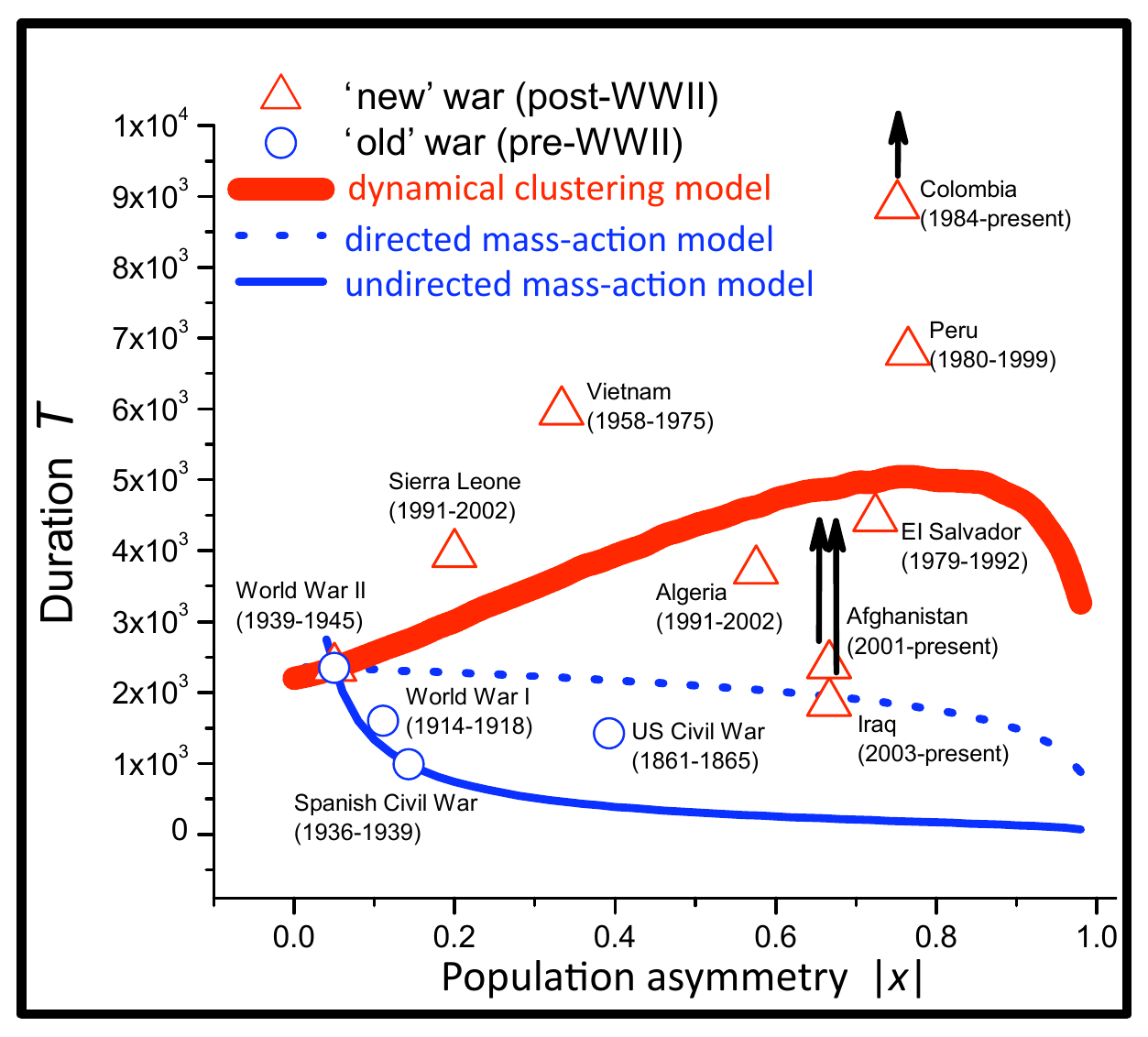}
\caption{(Color online) From Ref. \cite{duration}, our model gives an estimate of the duration of a conflict. Duration $T$ of human conflicts, as function of asymmetry $x$ between the two opposing populations $A$ and $B$. $x=|N_A(0)-N_B(0)|/[N_A(0)+N_B(0)]$. Data are up to the end of 2008, hence final datapoints for the three ongoing wars will lie above positions shown, as indicated by arrows. Lower two blue lines are the mass-action results.}
\label{fig1}
\end{figure}

Another criticism of Ref. \cite{nature} which has appeared in comments on the Internet, concerns the nature of the `information' that the cells have available to them in our generalized El Farol model for cell decision-making  \cite{nature}. As in the El Farol model itself, `information' here simply means something which acts as a common cue -- not necessarily a particular media source (e.g. CNN) or even type of media.  Indeed, we state in Ref. \cite{nature} that: ``Each group receives daily some common but limited information (for example, Éopposition troop movements, a specific religious holiday, even a shift in weather patterns). The actual content is relatively unimportant provided it becomes the primary input for the groupÕs decision-making process." This common information acts as a coordinating effect. Even if it is incorrect or inaccurate, it can still act to concentrate responses in a similar way. This crowding effect in strategy space is explained in detail, in the context of financial market burstiness, in Ref. \cite{book}.

So far we have focused on reproducing the statistical stylized facts for insurgent conflicts and terrorism. But given the development of a minimal model as described above and in Fig. 7 and 8, we can also ask the practical question: Can we estimate how long a conflict will last?
To address this in a simple way, we make the assumption of applying the law of pure attrition -- or more precisely, that a conflict lasts as long as it takes to reduce one side to a certain level of strength $N(t)$. The result is shown in Fig. 9 and the details are given in Ref. \cite{duration}. Our results show that a minority Red population experiences a longer survival time against a majority Blue force, than it would in the case of two equally balanced populations. This result is irrespective of whether the majority population adopts such internal grouping or not. Adding a third population with pre-defined group sizes allows the duration to be tailored. As shown, our findings compare favorably to real-world observations. We stress that these findings are not a simple consequence of either dilution leading to reaction slow-down, or of the specific cluster selection scheme that we chose. In our model, as in nature, opposing predator groups actively seek each other out at each timestep, even if their density is low, making this unlike simple chemical dilution, and hence unlike simple mass-action equations. Instead, our results emerge from the interplay between population asymmetry, the presence of clustering, and the intentional engagement between the two opposing populations. Although the specific consequences may vary by application area, we believe that related phenomena lying beyond mass-action predictions will arise in a wide range of physical, chemical, biological and social systems, whenever intra-population clustering coexists with inter-population reactions.

\begin{figure}
\includegraphics[width=0.8\linewidth]{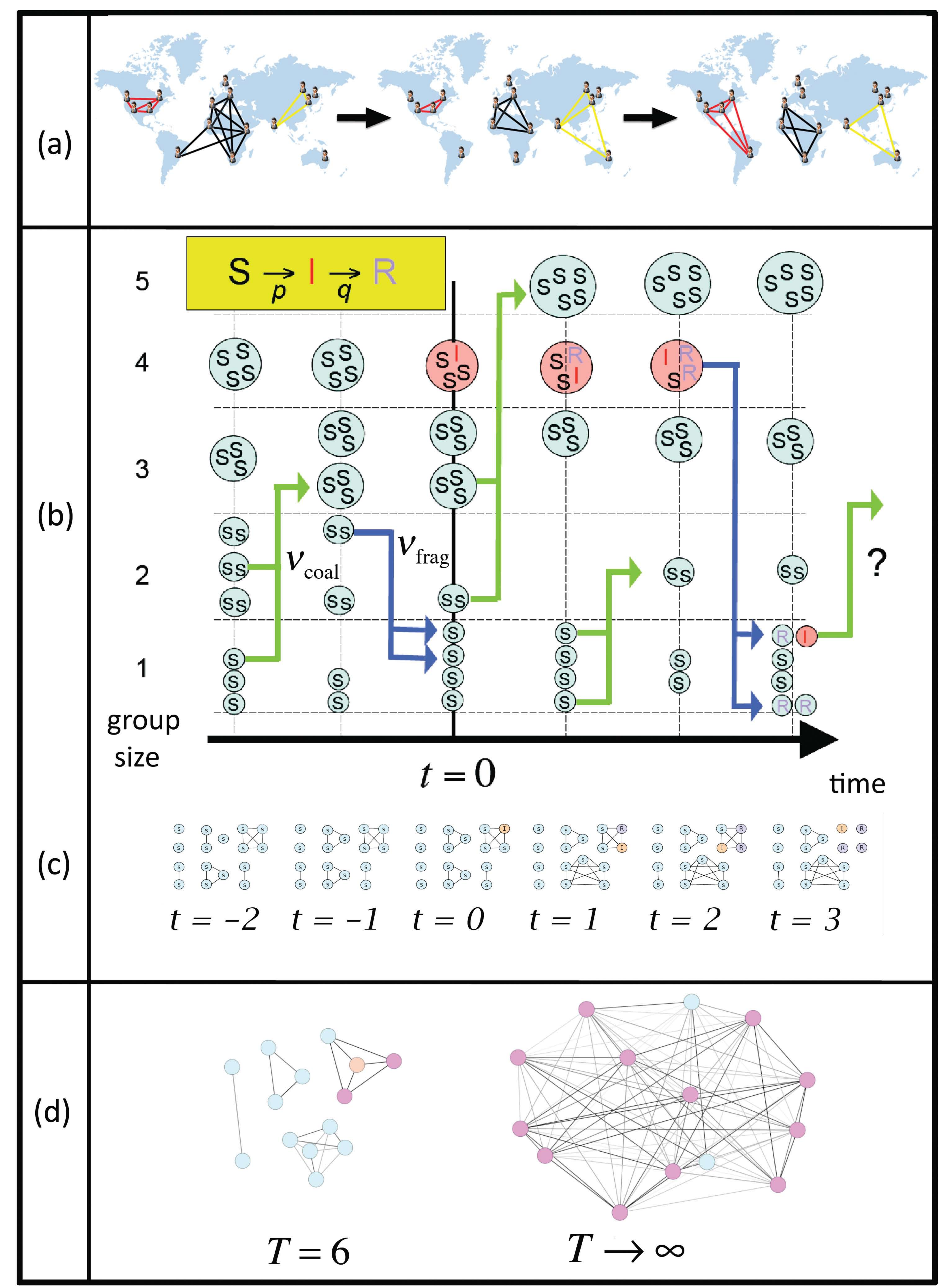}
\caption{(Color online) From Ref. \cite{ezsir}, the process of transmission through a population comprising a loose soup of cells (groups) which follow our model's simple dynamical rules for growth and break-up. a: Schematic of dynamical grouping of insurgents on the Internet, or Facebook etc.
b: Schematic of our model, featuring spreading in the presence of dynamical grouping via coalescence and fragmentation. Vertical axis shows number of cells (groups) of a given size at time $t$. c: Instantaneous network from Fig. 10b at each timestep.
d: Weighted network obtained by aggregating links over time-window $T$.}
\label{fig1}
\end{figure}

\section{SPREADING RUMORS AND MEMES IN AN INSURGENT POPULATION}
We now turn to look at the effect of spreading of a meme, or idea, or doctrine, or knowledge, within the Red population -- in order to understand how such populations might be persuaded and even infiltrated. This seems similar to an epidemic modeling problem, and hence one might be tempted to use one of the many approaches already developed. However, just as for models of conflict discussed earlier, dynamical models of spreading tend to fall into one of two extremes \cite{Koopman}, neither of which is particularly realistic for modern insurgent wars or terrorist activity -- and even less so for online cyberterrorism and cyberattacks. At one extreme, they assume that the social mixing dynamics are much faster than the spreading process, and hence that mass-action models can be adopted in which a continuum approximation can be used to generate differential equations based on calculus. This tends to be the extreme favored by mathematicians and physicists and engineers, since it unlocks the power of calculus and the vast spectrum of known properties of differential operators. At the other extreme, is the limit which has been favored by the social science community, in which the heterogeneity of social links is retained at the expense of assuming that the social network is static on the timescale of spreading-related events. 
Many state-of-the-art descriptions of viral transmission processes in real populations
incorporate system-specific details and considerations (e.g.
spatial topology, differential susceptibility) \cite{Keeling,May,cvespignani,schwartz,blasius}.
As indicated above, some of these focus on the well-mixed (i.e. mass-action) limit,
some of these focus on the limit of heterogeneous
networks -- and
some attempt to move between the two by adding patch-like
structure to mass-action models, or dynamical link rewirings to
network models \cite{schwartz,blasius}.

Our model (Fig. 8) in which dynamical regroupings happen spontaneously and sporadically over time, allows us to focus on the spreading dynamics in the realistic but  less well understood 
regime where the group-level dynamics and individual-level
transmission processes can evolve on the same timescale, and hence
the number and identity of a given individual's contacts can
change abruptly at any given moment in time. This is shown in Fig. 10 (see, for example, (a) and (b)) and is explained in detail in Ref. \cite{ezsir}. Most importantly, the
dynamical processes of social group or cell  formation/break-up and
person-to-person transmission of information, can co-exist on comparable timescales. We adopt the simple one-population form of our model (Appendix A). By varying the
probabilities of group coalescence ($\nu_{\rm coal}$) and
fragmentation ($\nu_{\rm frag}$) relative to the standard SIR
(Susceptible$\rightarrow$Infected$\rightarrow$Recovered)
probabilities for person-to-person
transmission ($p$) and individual recovery ($q$), the entire range
of relative timescales can be easily explored -- from a very
slowly changing insurgent cell structure (i.e. essentially a
static network with infrequent rewirings) through to a rapidly
changing cell structure (i.e. essentially a well-mixed
population). Our model only has four stochastic
parameters for the probabilities (and hence timescales) of the
individual level transmission and cell dynamics, i.e. $p$, $q$
for the SIR process, and $\nu_{\rm coal}$ and $\nu_{\rm frag}$
which describe the probability of cells coalescing or
fragmenting (Appendix A). Reference \cite{ezsir} shows explicitly that it reproduces the qualitative shapes of a wide range of empirical profiles associated with social, financial and biological spreading,
simply by varying these relative timescales. 
One implication of our findings is that conventional intelligence approaches in which the connections and nodes are sought assuming some quasi-static network, are likely to be unreliable at best -- and dangerously wrong at worst -- leading to misplaced analyses and operations and possibly ultimately endangering Blue personnel. Fuller details of these dynamical spreading results are given in Ref. \cite{ezsir}.

\section{OUTLOOK}
Our modeling approach described here, is characterized by two stages: First, our broad-brush dynamical Red Queen theory which describes the escalation between Red and Blue \cite{science}. This theory and analysis does not depend on the precise mechanism which changes the Red Queen's lead at any one time. Second, we provide a plausible microsocopic mechanistic model which captures more of the complexity shown in Fig. 1, with interacting populations comprising dynamically evolving cells in some loose and sporadically-changing structure. This model accounts for both the stylized facts of the timing of events and their severity. The parameters in our model are relatively few, and the model itself allows a range of analytic mathematical analysis to be performed for both the severity \cite{blazej} and the timing \cite{book}. Although simple and intuitive, the mechanisms incorporated in the model mimic certain real-world human behavioral features, such as (i) human decision-making under conditions of limited endogenous and exogenous information, (ii) the fragile and transitory nature of criminal groups, (iii) confidence levels, and desire for success, (iv) punishment and reward \cite{mike}  via the way in which strategy scores are updated, and (v) the tendency for human insurgents to occasionally coordinate actions using modern technology for long-distance communications, e.g. mobile phone calls and Internet use. Adding in the aspect of character to these models, which we are currently doing, will extend this work beyond the current stage where individuals (e.g. insurgents) are just particles, and will  allow us to examine human social, cultural and behavioral issues such as reward-punishment payoffs for cell members with different social and cultural backgrounds \cite{mike}. 

It may well turn out that other explanations of the stylized facts of human insurgency are possible. In fact, we both hope and expect this will happen in the near future -- just as the study of financial markets has spurred the fledgling field of Econophysics to become so productive over the past decade \cite{econophysics}. As more stylized facts become available, the competing theories can be judged against these benchmarks. It may also occur that, as in the study of financial markets, certain types of stochastic time-series-generating process can also reproduce the observed statistical features -- however, just as in the financial market field, it is well recognized that no deep understanding of market dynamics is offered by such models, other than the ability to replicate similar statistical patterns. By contrast, our goal is to deliver a model which is based on reasonable mechanisms of the dynamics of insurgencies at the cell level, with fairly minimal assumptions, and hence open up the path to a wide range of uses (e.g. scenario testing, evaluation of different strategies, interpretation of the ÔchangeÕ in a war through a surge etc.). 

We now comment on the comparison to cybergangs and street gangs. We found that when we analyzed the empirical distributions for Long Beach street gang sizes and online guild sizes for World of Warcraft \cite{gangs}, the empirical distributions were not power-law like. This can be explained by the fact that our data comprised  the actual membership of online guilds and gangs, as well as street gangs, as opposed to the number of objects who happen to be coordinated (e.g. online, or on the street) at any one time. The latter is likely to vary rapidly and spontaneously every day as members come online or onto the street, however the underlying membership would be expected to change more slowly over timescales of months. In addition, when individuals leave a street gang or an online guild, it is  unlikely that this happens because the entire gang or guild is disbanding -- hence the fragmentation process in our model is less realistic. Indeed, it is known that fission processes involving the partial dismantling of a large cell into just a few randomly chosen splinter-cells tends to generate non-power-law distributions, as is observed for street gangs and online guilds \cite{gangs}. In short, the rules for the coalescence-fragmentation in street gangs and guilds are likely to change when one considers longer-term membership, as compared to Figs. 7 and 8. In this case, we believe that the role of individual character will come more to the fore. This is indeed exactly what we found in Ref. \cite{ezsir} -- by adding character, in addition to some simple rules based on team formation, we found that we could reproduce the size distribution results for both cyber gangs and also street gangs on the monthly scale.  

This study could be opened up to other forms of collective human predation, such as the sinister threat from online child pornography rings. Even non-human predation can be considered, such as `battles' involving populations of pathogens within the immune system, or even the analogy to insurgency where parts of the immune system attack itself -- and where normal cells turn cancerous, generating primary tumors as well as secondary spreading through metastasis \cite{charley}. One particular example for which there is a wealth of data for collective human predation, is a financial market. In the fast, high-frequency regime of intraday trading, predatory algorithms can dominate the market at particular instances. Furthermore, they 
operate across multiple markets on the scale of hundreds, or event tens, of milliseconds, without regard for geographical boundaries. This connection between markets and predation may run even deeper, given the fact that many causes and drivers of social unrest may ultimately be linked to individual wealth and hence to the dynamics of the markets. Indeed, the lead article on the front page of the New York Times on Friday, July 18, 2008, featured what looked like a typical picture of insurgent activity, but noted below that the cause was actually the successive plunges in the Pakistan stock market over a two week period. This new area of coupled societal risks represents a huge future modeling challenge.

\section{ACKNOWLEDGEMENT} I am extremely grateful to the many collaborators that have made these works possible, including Mike Spagat, Brian Tivnan, Pak Ming Hui, Spencer Carran, Juan Camilo Bohorquez, Roberto Zarama, Amith Ravindar, Juan Pablo Calderon, Guannan Zhao, Elvira Restrepo and all other co-authors on the cited papers. I also gratefully acknowledge a grant from the Office of Naval Research (ONR): N000141110451. The views and conclusions contained in this paper
are those of the author and should not be interpreted
as representing the official policies, either expressed or implied, of any of the above named organizations,
to include the U.S. government. Subsequent draft revisions of this paper will be posted at the same arXiV e-print address, when updates of empirical analyses or model advances are available.

\appendix

\section{}
Here we consider the basic version of our model, stripped down to a simple form with no decision-making, and only one population -- the Red insurgency. Instead of having cells fragment when interacting with Blue, or when sensing imminent danger, we simply assign a  probability for them to fragment. The resulting model yields an exponentially cutoff 2.5-exponent power-law for the distribution of cell sizes. We note that generalizations of this model have appeared in the literature -- in particular, Ref. \cite{blazej} contains a number of relevant generalizations, including a variable number of agents in time $N(t)$. A later paper provides a different derivation of the same basic result as the one below \cite{clauset2}, reaching similar conclusions to our earlier publication  (Ref. \cite{blazej}) concerning the remarkable robustness of the 2.5 exponent to variations in the model mechanisms.

Assuming that the civilian population is just some passive background that absorbs the strength of each cell when that cell acts, the distribution of civilian casualties should have a similar distribution to that of the insurgent cells (see main text for a fuller discussion of this point). Analysis of a simple version of this model was completed earlier by d'Hulst and Rodgers \cite{rodgers}, and real-world applications have focused on financial markets -- however the derivation below features general values $\nu_{\rm frag}$ and $\nu_{\rm coal}$. At each timestep, the internal coherence of a population of $N$ objects (which we refer to as an `agents' to acknowledge application to human and/or cyber systems) comprises a heterogenous soup of cells. Within each cell, the component objects have a strong intra-cell coherence. Between cells, the inter-cell coherence is weak. An agent $i$ is then picked at random -- or equivalently, a cell is randomly selected with probability proportional to size. Let $s_i$ be the size of the
cell to which this agent belongs. With probability $\nu_{\rm frag}$, the coherence of a given cell fragments completely into $s_i$ cells of size one. If it doesn't fragment, a second cell is randomly selected with probability again proportional to size -- or equivalently, another agent $j$ is picked at random. With probability $\nu_{\rm coal}$, the two cells then coalesce (or develop a common `coherence' in terms of their thinking or activities). As discussed in the main text, Kenney provides a wealth of case-study support for thinking of an insurgency as a loose soup of fragile cells  \cite{kenney}, as do Gambetta \cite{gambetta} and Robb \cite{robb}.

The Master Equation is as follows:
\begin{eqnarray}
\frac{\partial n_s}{\partial
t}&=&{\frac{\nu_{\rm coal}}{N^2}\sum^{s-1}_{k=1}kn_k(s-k)n_{s-k}}-{\frac{\nu_{\rm frag}sn_s}{N}}-{\frac{2\nu_{\rm coal}sn_s}{N^2}\sum^{\infty}_{k=1}kn_k}
\ , \quad s\geq2  \ , \label{eq:genez1}\\
 \frac{\partial n_1}{\partial t}&=&\frac{\nu_{\rm frag}}{N}\sum^{\infty}_{k=2}k^2n_k-\frac{2\nu_{\rm coal}n_1}{N^2}\sum^{\infty}_{k=1}kn_k \ . \label{eq:genez2}
\end{eqnarray}
Note here we make an approximation that $N\rightarrow\infty$. The
terms on the right hand side of Eq.~(\ref{eq:genez1}) represent
all the ways in which $n_s$ can change.
In the equilibrium state:
\begin{eqnarray}
sn_s&=&\frac{1-\nu_{\rm frag}}{(\nu_{\rm frag}+2\nu_{\rm coal})N}\sum^{s-1}_{k=1}kn_k(s-k)n_{s-k}
\ ,
\quad s\geq2 \ , \label{eq:ezeq}\\
n_1&=&\frac{\nu_{\rm frag}}{2(1-\nu_{\rm frag})}\sum^{\infty}_{k=2}k^2n_k\
\ . \label{eq:ezeq1}
\end{eqnarray}
Consider
\begin{equation}
G[y]=\sum^{\infty}_{k=0}kn_ky^k=n_1y+\sum^{\infty}_{k=2}kn_ky^k
\equiv n_1y+g[y] \ , \label{eq:generating}
\end{equation}
where $y$ is a parameter and g[y] governs the cell size
distribution $n_k$ for $k \geq 2$. Multiplying Eq.~(\ref{eq:ezeq})
by $y^s$ and then summing over $s$ from $2$ to $\infty$, yields:
\begin{equation}
g[y]=\frac{1-\nu_{\rm frag}}{(\nu_{\rm frag}+2\nu_{\rm coal})N}G[y] \ ,
\end{equation}
i.e.
\begin{equation}
g[y]^2-\left(\frac{\nu_{\rm frag}-2\nu_{\rm coal}}{\nu_{\rm coal}}N-2n_1y\right)g[y]+n_1^2y^2=0\label{eq:g[y]}
\ .
\end{equation}
From Eq.~(\ref{eq:generating}), $g[1]=G[1]-n_1$. Substituting this
into Eq.~(\ref{eq:g[y]}) and setting $y=1$, we can solve for
$g[1]$
\begin{equation}
g[1]=\frac{\nu_{\rm coal}}{\nu_{\rm frag}+2\nu_{\rm coal}}N \ .
\end{equation}
Hence
\begin{equation}
n_1=N-g[1]=\frac{\nu_{\rm frag}+\nu_{\rm coal}}{\nu_{\rm frag}+2\nu_{\rm coal}}N \
.
\end{equation}
Substituting this into Eq.~(\ref{eq:g[y]}) yields
\begin{equation}
g[y]^2-\left(\frac{\nu_{\rm frag}+2\nu_{\rm coal}}{\nu_{\rm coal}}N-\frac{2N(\nu_{\rm frag}+\nu_{\rm coal})}{\nu_{\rm frag}+2\nu_{\rm coal}}y\right)g[y]+\frac{(N(\nu_{\rm frag}+\nu_{\rm coal}))^2}{(\nu_{\rm frag}+2\nu_{\rm coal})^2}y^2=0
\ .
\end{equation}
We can solve this quadratic for $g[y]$
\begin{equation}
g[y]=\frac{(\nu_{\rm frag}+2\nu_{\rm coal})N}{4\nu_{\rm coal}}\left(2-\frac{4(\nu_{\rm frag}+\nu_{\rm coal})\nu_{\rm coal}}{(\nu_{\rm frag}+2\nu_{\rm coal})^2}y-2\sqrt{1-\frac{4(\nu_{\rm frag}+\nu_{\rm coal})\nu_{\rm coal}}{(\nu_{\rm frag}+2\nu_{\rm frag})^2}y}\right)
\ ,
\end{equation}
which can be easily expanded
\begin{equation}
g[y]=\frac{(\nu_{\rm frag}+2\nu_{\rm coal})N}{2\nu_{\rm coal}}\sum^{\infty}_{k=2}\frac{(2k-3)!!}{(2k)!!}\left(\frac{4(\nu_{\rm frag}+\nu_{\rm coal})\nu_{\rm coal}}{(\nu_{\rm frag}+2\nu_{\rm coal})^2}y\right)^k
\ .
\end{equation}
Comparing with the definition of $g[y]$ in
Eq.~(\ref{eq:generating}) shows that
\begin{equation}
n_s=\frac{\nu_{\rm frag}+2\nu_{\rm coal}}{2\nu_{\rm coal}}\frac{(2s-3)!!}{s(2s)!!}\left(\frac{4(\nu_{\rm frag}+\nu_{\rm coal})\nu_{\rm coal}}{(\nu_{\rm frag}+2\nu_{\rm coal})^2}\right)^s
\ .
\end{equation}
We now employ Stirling's series
\begin{equation}
ln[s!]=\frac{1}{2}ln[2\pi]+\left(s+\frac{1}{2}\right)ln[s]-s+\frac{1}{12s}-...
\ .
\end{equation}
Hence for $s\geq2$, we find
\begin{equation}
n_s\approx\left(\frac{(\nu_{\rm frag}+2\nu_{\rm coal})e^2}{2^{3/2}\sqrt{2\pi}\nu_{\rm coal}}\right)\left(\frac{4(\nu_{\rm frag}+\nu_{\rm coal})\nu_{\rm coal}}{(\nu_{\rm frag}+2\nu_{\rm coal})^2}\right)^s\frac{(s-1)^{2s-3/2}}{s^{2s+1}}N
\ ,
\end{equation}
which implies that
\begin{equation}
n_s \sim
\left(\frac{\nu_{\rm coal}^{s-1}(\nu_{\rm frag}+\nu_{\rm coal})^s}{(\nu_{\rm frag}+2\nu_{\rm coal})^{2s-1}}\right)s^{-5/2}\ \ . \label{eq:power}
\end{equation}
In the limit $s\gg 1$, this is formally equivalent to saying that
\begin{equation}
n_s \sim
{\rm exp}(-s/s_0) s^{-5/2}\label{eq:power2}
\end{equation}
where
\begin{equation}
s_0=-\left[{\rm ln} \left(\frac{4(\nu_{\rm frag}+\nu_{\rm coal})\nu_{\rm coal}}{(\nu_{\rm frag}+2\nu_{\rm coal})^2}\right)\right]^{-1}\ \ .
\end{equation}
For large cell sizes (i.e. large $s$ such that $s\sim O(N)$) the power law behaviour is masked by the
exponential function. The equilibrium state for the distribution
of cell sizes can therefore be considered a power-law with
exponent $\alpha\sim5/2=2.5$, together with an exponential cut-off.
In the human context, the fact that the interactions are effectively distance-independent as far as Eq. (A1) is concerned, captures the fact that we wish to model systems where messages can be transmitted over arbitrary distances (e.g. modern human communications). Bird calls and chimpanzee interactions in complex tree canopy structures can also mimic this setup, as may the increasingly longer-range awareness that arises in larger animal, fish, bird and insect groups. In a human/biological context, a justification for choosing a cell with a probability which is proportional to its size, is as follows: a cell with more members has more chances of initiating an event. It will also be more likely to find members of another cell more frequently, and hence be able to synchronize with them -- thereby synchronizing the two cells. 
It is well documented that cells of living objects (e.g. animals, people) may suddenly scatter in all directions (i.e. complete fragmentation as in Eq. (A1)) when its members sense danger, simply out of fear or in order to confuse a predator. Such fleeing behavior was discussed at length in the classic 1970 work `Protean Defence by Prey Animals' by D. A. Humphries and P.M. Driver, Oecologia (Berl.) {\bf 5}, 285-302 (1970). 

This model also offers an alternative explanation for a variety of other complex phenomena which have been found to exhibit a robust 2.5 power-law. Gabaix {\it et al.} \cite{Gabaix} found a common power-law distribution for individual transaction sizes with $\alpha=2.5\pm 0.1$, for the London Stock Exchange, the NYSE, and the Paris Bourse. Interpreting $N$ as the average aggregate demand for stocks, this demand $N$ gets shaped into a distribution of demand `clusters' representing potential orders of a given size $s$. 
Since it is reasonable to expect orders to be realized at random, the distribution of individual transaction sizes is proportional to the distribution of clusters of potential orders -- hence $\alpha = 2.5$. Similarly, Richardson \cite{richardson} concluded that the distribution of approximately $10^3$ gangs in Chicago, and in Manchoukuo in 1935, separately followed a truncated power-law with $\alpha \approx 2.3$. Interpreting $N$ as the number of potential gang members in each case, with each comprising a transient soup of clusters which tend to combine or fragment over time, yields $\alpha = 2.5$. In a similar way, the robust time-dependence of a power-law with $\alpha\approx 2.4$ in a recent New York garment industry study \cite{felix} can be reinterpreted as a repartitioning of trading interactions, with multi-component clusters continually being built up as part of common jobs (i.e. coalescence) and then dissolving upon completion (i.e. fragmentation). 
For collections of $N$ neurons \cite{chialvo}, we can imagine a dynamical coalescence-fragmentation grouping process in which groups of neurons become synchronized, and then this synchronization ultimately fragments. (Members of the same group need not be physically adjacent to each other). When an entire group fires, it creates a measurable activity equal to \cite{chialvo} the group size $s$. Hence the resulting activity distribution will follow a new power-law given by $s\times p(s)$. The resulting power-law exponent $(\alpha-1) = 1.5$, which is exactly the famous empirical $3/2$ value \cite{chialvo}. 
We note that although competing theories exist for many of these applications (i.e. Chialvo \cite{chialvo} for neural dynamics and Gabaix {\it et al.} \cite{Gabaix} for markets), we know of no other single mechanism which is simultaneously physically plausible for each application area {\em and} which can also explain a mysterious recent finding in the field of superconductivity \cite{aip}.

\end{document}